\documentclass[reprint, prl, amsmath, amssymb, aps, superscriptaddress, showpacs, longbibliography]{revtex4-1}

\usepackage{graphicx, color}
\usepackage[normalem]{ulem}

\usepackage[colorlinks=true]{hyperref}
\hypersetup{pdfpagemode=None, linkcolor=black, citecolor=blue, urlcolor=blue}
\usepackage{orcidlink}

\usepackage[T1]{fontenc}
\usepackage[utf8]{inputenc}

\usepackage{amssymb}
\usepackage{amsmath}
\usepackage{commath}
\usepackage{graphicx,bm}
\usepackage{verbatim}
\usepackage{ulem}

\usepackage[textsize=scriptsize, colorinlistoftodos]{todonotes}
\usepackage{xcolor}

\definecolor{britishracinggreen}{rgb}{0.0, 0.26, 0.15}
\definecolor{bulgarianrose}{rgb}{0.28, 0.02, 0.03}
\definecolor{darkred}{rgb}{0.90,0,0}
\definecolor{darkgreen}{rgb}{0,0.45,.15}
\definecolor{darkblue}{rgb}{0,0,1}
\definecolor{orange}{cmyk}{0,0.6,0.8,0}
\definecolor{lightblue}{rgb}{0.3,0.5,1}
\definecolor{lightgreen}{rgb}{0.4,0.80,.4}

\newcommand{\ssec}[1]{\noindent\emph{#1}\,---}

\newcommand{\kF}{k_F}
\newcommand{\eF}{\varepsilon_F}

\newcommand{\Rmid}{R_{\rm mid}}
\newcommand{\nnod}{n_{\rm nodal}}

\newcommand{\Ntot}{N_\mathrm{tot}}

\begin{document}
\title{Rotation-induced supersolidity in a spin-imbalanced unitary Fermi gas}

\author{Bu\u{g}ra T\"uzemen\,\orcidlink{0000-0002-8619-754X}}\email[Contact author: ]{btuzemen@ifpan.edu.pl}
\affiliation{Institute of Physics, Polish Academy of Sciences, Al. Lotnik\'{o}w 32/46, 02-668 Warsaw, Poland}

\author{Gabriel Wlaz\l{}owski\,\orcidlink{0000-0002-7726-5328}}
\affiliation{Faculty of Physics, Warsaw University of Technology, Ulica Koszykowa 75, 00-662 Warsaw, Poland}
\affiliation{Department of Physics, University of Washington, Seattle, Washington 98195--1560, USA}

\author{Piotr Magierski\,\orcidlink{0000-0001-8769-5017}}
\affiliation{Faculty of Physics, Warsaw University of Technology, Ulica Koszykowa 75, 00-662 Warsaw, Poland}
\affiliation{Department of Physics, University of Washington, Seattle, Washington 98195--1560, USA}

\author{Klejdja Xhani\,\orcidlink{0000-0003-0713-8523}}\email[Contact author: ]{klejdja.xhani@polito.it}
\affiliation{Department of Applied Science and Technology, DISAT, Politecnico di Torino, Italy}
\date{\today}

\begin{abstract}
We show that a supersolid can be spontaneously generated by a persistent current in a spin-imbalanced unitary Fermi gas in a ring. As polarization increases, stable superflow turns dissipative via vortex emission before superfluid and crystalline order coexist. Pairing and spin densities are modulated with an even number of nodes, where the pairing phase jumps by $\pi$ and excess majority spins accumulate. Even at fixed polarization, the modulation period changes with the winding number, unlike in FFLO states. The modulation slides freely, supporting a gapless translational Goldstone mode.
\end{abstract}

\maketitle
\ssec{Introduction.}
Spin imbalance provides a powerful route to unconventional pairing in fermionic superfluids, where competition between pairing and polarization can destabilize the standard BCS order in favor of inhomogeneous phases. The Fulde-Ferrell-Larkin-Ovchinnikov (FFLO) state is the canonical example, characterized by a spatially modulated order parameter whose wavevector is set by the mismatch between the Fermi surfaces of the two spin species~\cite{Fulde_1964, Larkin_1965}. While these phases are of broad interest across condensed matter, nuclear, and atomic physics~\cite{Buzdin_2005, Sedrakian_2006, Zwierlein_2006, Partridge_2006, Stein_2014, Torma_2018, Pirolo_2026}, direct observation remains elusive, with only indirect signatures reported in organic superconductors~\cite{Mayaffre_2014} and 1D gases~\cite{Liao_2010}. Three dimensions are particularly challenging: the phase is structurally fragile, susceptible to fluctuations, and tends to drive phase separation rather than a robust modulated condensate~\cite{Radzihovsky_2009, Radzihovsky_2010, Radzihovsky_2011}. At unitarity, density-functional theory expands the Larkin-Ovchinnikov (LO) regime~\cite{Bulgac_2008}, though only under explicit structural constraints. Unconstrained simulations instead reveal disordered droplets or irregular nodes~\cite{Magierski_2019, Tuzemen_2023}, demonstrating that 3D inhomogeneous order is exceptionally sensitive to geometry, topology, and initial conditions.

The realization of a fermionic supersolid in which the pairing field itself carries the crystalline order remains an elusive goal, whereas bosonic supersolids are well established: two Goldstone modes are associated with the broken continuous gauge and translational symmetries, and a Higgs mode accounts for the amplitude sector of the spectrum~\cite{Leonard_2017, Li_2017, Bottcher_2019, Chomaz_2019, TanziPRL_2019, Norcia_2021, Recati_2023, Guo_2019, Tanzi_2019, Natale_2019, Leonard_2017b, Hertkorn_2021, Geier_2021, Biagioni_2024, Tanzi_2021, Roccuzzo_2020, Roccuzzo_2022, Trabucco_2026}. A toroidal trap provides a natural setting to bridge this gap.

Its multiply connected geometry supports persistent currents with a quantized winding number $w_0$~\cite{Ramanathan_2011, DelPace_2022, Xhani_2025}. Imprinting such a current gives every atom a momentum $\hbar w_0/(2R)$, with $R$ the ring radius. This momentum is set by a topological integer, independently of the Fermi-surface mismatch that sets the FFLO modulation. Rotation can even drive the superfluid–supersolid transition in dipolar condensates~\cite{Schubert2026}, and the interplay between quantized circulation and a density-modulated order parameter has recently been modeled for annular dipolar supersolids~\cite{Tengstrand_2023, Mukherjee_2025, Preti_2026}. Angular FFLO states have likewise been studied in toroidal traps and rings~\cite{Yanase_2009, Yoshida_2011, Yoshida_2012}, where rotation can shift the nodal structure, but these were static calculations, and the base wavevector remains set by the Fermi-surface mismatch. Quantized circulation introduces an additional selection mechanism: at fixed polarization, the winding number can tune the modulation wavevector of the order parameter.

In this Letter, we realize this concept by tracking the unconstrained dynamics of a spin-imbalanced strongly interacting unitary Fermi gas prepared in a toroidal trap with winding number $w_0$. Without external driving or imposed spatial symmetry, the persistent flow either survives or, depending on polarization and $w_0$, breaks down through vortex emission, as in the unpolarized gas~\cite{Xhani_2025, Tuzemen_2026}, or spontaneous supersolid ordering. In the supersolid phase, $w_0$ selects the modulation period of a joint pairing and spin-density pattern in addition to and distinct from the Fermi-surface mismatch. Each nodal line acts as a $\pi$ junction within a globally coherent condensate. Once the nodal crystal has formed, the pattern slides freely and supports a linear translational mode consistent with a gapless Goldstone branch.

\begin{figure*}[t]
\centering
\includegraphics[trim={7pt 8pt 10pt 7pt}, clip, width=\textwidth]{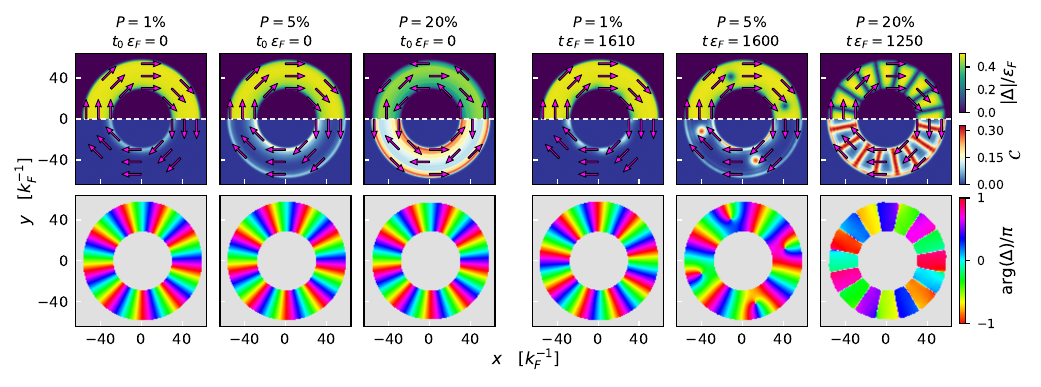}
\caption{Three dynamical regimes at fixed winding number $w_0=9$. Columns pair the initial state with the final state for polarizations $P=1\%$, $5\%$, and $20\%$. Upper row: pairing amplitude $|\Delta|/\eF$ (upper half of each disc) and spin-density contrast (lower half); arrows show the direction of the total current density $\mathbf{j}=\mathbf{j}_\uparrow+\mathbf{j}_\downarrow$. Lower row: phase $\arg(\Delta)/\pi$. At $P=1\%$, the persistent current survives. At $P=5\%$, the inner edge emits polarized vortices, causing local $2\pi$ phase slippage. At $P=20\%$, the pairing field develops a periodic nodal structure with $\nnod=16$ domains whose relative phase is close to $\pi$.}
\label{fig:regimes}
\end{figure*}

\ssec{Emergence of the supersolid regime.}
We begin by surveying the dynamics over polarization $P=\frac{N_{\uparrow}-N_{\downarrow}}{N_{\uparrow}+N_{\downarrow}}$ and initial winding number $w_0$, seeking regimes in which persistent flow can generate spatially ordered superfluid states. Our aim is not to map the full phase diagram in the $(P,w_0)$ plane, but rather to identify a regime where supersolid behavior emerges and examine its properties in detail. We employ a quasiparticle-resolved density-functional approach~\cite{Bulgac_2012} whose equations have Bogoliubov–de Gennes form. The density, pairing field, currents, and other observables are obtained self-consistently from the evolving quasiparticle orbitals, without imposing a modulation or spatial symmetry. We use units in which $\hbar$ and the atomic mass are set to one, and express lengths, energies, velocities, and times through the Fermi wavevector $\kF=(3\pi^2\rho_0)^{1/3}$ and the Fermi energy, where $\rho_0$ is the total density in the bulk of the ring. The gas is confined to an annulus with inner and outer radii $R_{\rm in}=30\,\kF^{-1}$ and $R_{\rm out}=56\,\kF^{-1}$ and is uniform along the ring axis $z$. Details of the formalism and numerical implementation are given in the End Matter.

The scan proceeds in two steps: (i) we prepare a spin-imbalanced gas in a toroidal trap with an imprinted persistent current of winding $w_0$; (ii) we evolve the system and analyze the structures that emerge. The dynamical destabilization of the system stems from two key mechanisms: pair-breaking gap suppression~\cite{Xhani_2025, Tuzemen_2026} and counterflows similar to those carried by spin-polarized Andreev states in polarized vortex cores~\cite{Magierski_2022}. Both effects combine to drive the breakdown of persistent flow. Our focus here is on the state produced by the instability. As shown in Fig.~\ref{fig:regimes}, three distinct regimes emerge as polarization increases. At $P=1\%$, the current remains stable. At $P=5\%$, the instability is relieved through vortex emission, as in the balanced gas of Refs.~\cite{Xhani_2025, Tuzemen_2026}; here, however, the vortices harbor the excess majority particles in their cores. At $P=20\%$, relaxation follows a fundamentally different route. The pairing field develops a periodic azimuthal modulation with $\nnod=16$ nodal lines where $\vert{}\Delta\vert{}$ drops sharply and the spin-density contrast, $\mathcal{C} = \vert{}\rho_\uparrow - \rho_\downarrow\vert{} / \rho_0$, is strongly enhanced. The pattern forms spontaneously by $t\,\eF\approx150$ and persists throughout the evolution. As in the $P=5\%$ case, the instability starts at the inner edge, but now dips in $\vert{}\Delta\vert{}$ extend across the bulk into nodal lines rather than detaching as vortices. These nodal lines are stabilized by Andreev bound states~\cite{Magierski_2019, Tuzemen_2020,Magierski_2021}, ultimately producing a state with well-defined spatial periodicity which is not imposed externally but emerges spontaneously while the system remains superfluid: a defining feature of supersolidity~\cite{Andreev_1969, Chester_1970, Leggett_1970, Boninsegni_2012}. In the following, we demonstrate that the resulting state exhibits supersolid properties.

\begin{figure}[t]
\centering
\includegraphics[width=\columnwidth]{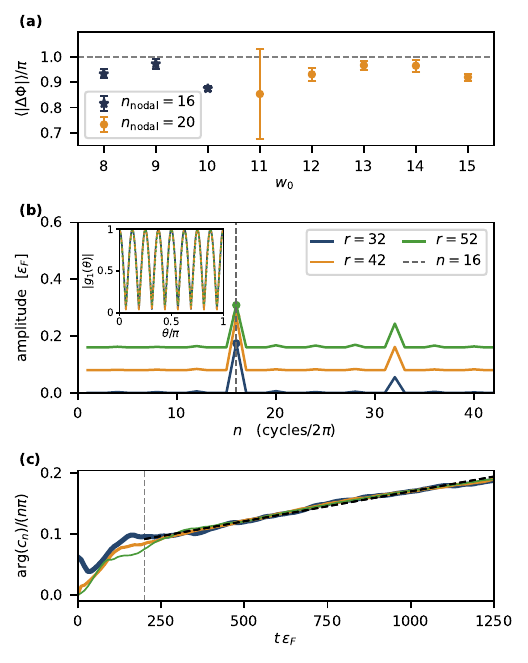}
\caption{Structure of the modulated state at $P=20\%$. (a) Mean phase jump per nodal line, $\langle|\Delta\Phi|\rangle/\pi$, against $w_0$, measured on the ring at $\Rmid = 42\kF^{-1}$ at the final time. The marker shape encodes the nodal count: $\nnod=16$ (stars) and $\nnod=20$ (circles). Error bars give the standard deviation over the nodes. (b) Amplitude spectrum of $|\Delta(\theta)|$ against angular wavenumber $n$ for $w_0=9$ at $t\,\eF=1250$, at $r=32$, $42$ and $52\,\kF^{-1}$, offset vertically for clarity. The peak sits at $\nnod=16$ at every radius. Inset: phase coherence $|g_1(\theta)|$ along the ring at the same radii; it returns to unity at every nodal spacing $2\pi/\nnod$. (c) Collective angular coordinate of the pattern, $\theta_0=\arg(c_{\nnod})/\nnod$, in units of $\pi$. Dashed lines are a joint linear fit for $t\eF\geq200$ (vertical line), with one slope at all three radii, $\omega_c=3.066\times10^{-4}\,\eF$.}
\label{fig:order}
\end{figure}

\ssec{Pairing and crystalline order.}
To map out the internal structure of the modulated state, we first examine the phase structure for initial winding numbers $w_0 = 8\text{--}15$ [Fig.~\ref{fig:order}(a)]; for $w_0<8$, no periodic structure forms. At each nodal line, the order parameter undergoes a rapid phase flip, acting as a $\pi$-junction reminiscent of the nodal planes in the LO state~\cite{Radzihovsky_2009, Radzihovsky_2011}, so that the ring forms a $\pi$-junction analog of a Josephson-junction necklace~\cite{Pezze2024}. Across all examined winding numbers, the mean phase difference $\langle\vert{}\Delta\Phi\vert{}\rangle$ remains bounded between $0.85\pi$ and $0.97\pi$. We focus on $w_0=9$ at $P=20\%$ as our primary showcase throughout: its mean phase difference comes closest to $\pi$ ($\approx 0.97\pi$) [Fig.~\ref{fig:order}(a)], and among states with comparably well-defined $\pi$ junctions, it carries the smallest $w_0$. This state sits far from the stability limit at $w_0=16$, where the modulated state is no longer stable, so its crystal forms with minimal transient disruption. Crucially, this robust phase-step consistency confirms the global gauge order. Uncoupled, independent condensates would instead yield uncorrelated relative phases, whereas the small node-to-node standard deviations [in Fig.~\ref{fig:order}(a)] show that the phase differences remain tightly restricted around their mean values. The periodic modulation thus resides on a pairing field that remains phase-coherent around the entire circumference.

The bulk nature and phase coherence of this crystalline order are established in Fig.~\ref{fig:order}(b) for $w_0=9$. The Fourier amplitude spectrum of $\vert{}\Delta(\theta)\vert{}$, where $\theta$ denotes the azimuthal angle in polar coordinates, displays a sharp fundamental peak at angular wavenumber $\nnod=16$. Throughout, we define $\nnod$ as the position of this strongest peak in the fast Fourier transform of $\vert{}\Delta(\theta)\vert{}$ along the ring (see End Matter). This peak position remains identical across all radial slices from $r=32$ to $52\,\kF^{-1}$, which shows that the lattice belongs to the bulk condensate rather than the ring edges. The inset of Fig.~\ref{fig:order}(b) shows the first-order coherence of the pairing field along the ring,
\begin{equation*}
g_1(r,\theta)=\frac{\int_0^{2\pi}d\theta'\,\Delta^*(r,\theta')\,\Delta(r,\theta'+\theta)}{\int_0^{2\pi}d\theta'\,\vert{}\Delta(r,\theta')\vert{}^2}.
\end{equation*}
Its modulus returns to unity ($\vert{}g_1\vert{}>0.99$) at the nodal spacing $\theta=2\pi/\nnod$ at all three radii. Shifted by one period, the pairing field therefore repeats itself, so every domain is locked to its neighbor by the same phase difference. Together with the mean phase differences in Fig.~\ref{fig:order}(a), this spatial phase locking provides direct evidence for global gauge order. While the time-dependent density functional assigns a definite local phase to the pairing field [Eqs.~(\ref{eq:aslda})--(\ref{eq:tdaslda})], it does not enforce phase agreement across distinct junctions. The resulting phase alignment across all sixteen domains is thus a purely emergent outcome of the dynamics. Quantized rotation acts primarily during pattern formation: once the nodal lines appear, the phase winding is no longer defined, yet the crystal retains the period selected by $w_0$. Because the initial imprint fixes only total circulation without specifying node coordinates, the pattern spontaneously breaks continuous translational symmetry. Finally, the modulation is visible in the density, as reflected in the spin-density contrast, and in the pairing amplitude, demonstrating the coexistence of superfluid and crystalline order.

Two structural rules govern the observed period. First, because the pairing field changes sign across each nodal line, single-valuedness around the ring requires an even number of nodes. Second, the period depends on both the initial circulation and the polarization. At fixed $P=20\%$, increasing $w_0$ from $8\text{--}10$ to $11\text{--}15$ raises the nodal count from $\nnod=16$ to $20$ [marker shapes in Fig.~\ref{fig:order}(a)], while the Fermi-surface mismatch stays at $\Delta k_F=k_{F\uparrow}-k_{F\downarrow}=0.13\,\kF$. Since $\nnod$ is the same at every radius [Fig.~\ref{fig:order}(b)], the corresponding wavevector $q_\Delta=\nnod/(2r)$ is not a single number: for $\nnod=16$ it decreases from $0.25\,\kF$ at $r=32\,\kF^{-1}$ to $0.15\,\kF$ at $r=52\,\kF^{-1}$. The period is thus fixed in angle by $\nnod$ rather than in length by $\Delta k_F$. This distinguishes the observed modulation from the conventional LO mechanism, in which a single pairing wavevector is set by the Fermi-surface mismatch: the mismatch remains relevant, but the winding number provides an additional control over the selected period.

A notable distinction from the dipolar Bose supersolids commonly studied experimentally is the structure connecting neighboring crystalline domains. In the coherent Bose-supersolid regime, density maxima remain coupled through a low-density superfluid background~\cite{Recati_2023, Biagioni_2024}. Here, instead, the pairing field vanishes exactly along the nodal lines, $\Delta=0$, so that viewed solely through $\Delta$, the modulated state resembles an array of disconnected superfluid domains. The matter distribution, however, remains connected: the nodal regions retain a finite particle density and are populated predominantly by the excess majority component, as expected for spin-imbalanced nodal states of LO character~\cite{Bulgac_2008,Radzihovsky_2011}. The presence of these zeros also changes the meaning of circulation. For a nonsingular pair field, the winding is defined as $w=(2\pi)^{-1}\oint_{\mathcal C}\nabla\arg[\Delta]\cdot d\mathbf l$, but this quantity ceases to be well defined once the integration contour intersects a nodal line, where the phase of $\Delta$ is undefined. Thus, in the supersolid state the integer winding cannot in general be used to characterize the remaining superfluid motion.

Figure~\ref{fig:order}(c) tracks the motion of the periodic structure itself via the Fourier coefficient $c_{\nnod}(t) = (2\pi)^{-1}\int_0^{2\pi} \vert{}\Delta(\theta,t)\vert{} e^{i \nnod\theta} \, d\theta$, whose argument yields the collective angular coordinate $\theta_0 = \arg(c_{\nnod}) / \nnod$. A pattern that rotates rigidly at frequency $\omega_c$ satisfies $\vert{}\Delta(\theta,t)\vert{} = f(\theta - \omega_c t)$, which implies $c_{\nnod}(t) = e^{i \nnod\omega_c t} c_{\nnod}(0)$. Across all three radial slices, $\theta_0(t)$ increases linearly in time with a uniform drift velocity $\omega_c = 3.066\times10^{-4}\,\eF$. Continuous rotational symmetry makes the system's energy independent of the absolute node positions, so uniform translations cost zero energy. This symmetry-protected gapless origin supports the free sliding observed in panel (c).

\begin{figure}[t]
\centering
\includegraphics[width=\columnwidth]{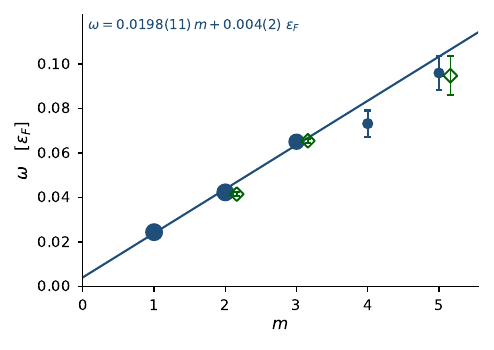}
\caption{Translational Goldstone branch at $P=20\%$, $w_0=9$. Filled circles: kick response frequency at wavenumber $m$; open diamonds: kicks at $m=11$, $13$, and $14$, plotted at reduced wavenumber $\nnod-m$. Circle size scales with spectral power. Solid line: weighted fit to $m=1$--$5$, $\omega/\eF=0.0198(11)\,m+0.004(2)$.}
\label{fig:goldstone}
\end{figure}

\ssec{Translational Goldstone branch.}
Spontaneous breaking of continuous translational symmetry gives rise to a low-energy excitation branch corresponding to vibrations of the crystal-like supersolid structure. Due to the geometry of the system, the excitation branch consists of discrete states corresponding to excitation of vibrational waves. We extract the spectrum by investigating the system's dynamical response to periodic perturbations of the nodal structure. Namely, we couple the perturbation to the spin-imbalance distribution through the potential $\delta V_\sigma = s_\sigma V_0 e^{-(t-t_k)^2/2t_w^2} \cos[m(\theta-\phi_0)]$, with $s_\downarrow = -s_\uparrow = 1$ and amplitude $V_0 = 0.01\,\eF$; $\phi_0$ is the angular position of a nodal line, and the pulse center $t_k$ and width $t_w$ are given in the End Matter. We evolve each perturbed state for a given $m$ alongside an unkicked control trajectory ($V_0=0$). Across all perturbed and unperturbed runs, the crystal preserves its $\nnod=16$ structure, confirming that we are in a weak perturbation limit.

To isolate the collective dynamics, we record the angular projection of the spin-density contrast, $F_m(r,t) = (2\pi)^{-1}\int d\theta \cos[m(\theta-\phi_0)] [\rho_\uparrow - \rho_\downarrow]$, at ten radial slices across the ring. Because the supersolid rotates uniformly, its initial configuration cannot serve as a static reference (as done, e.g., in Ref.~\cite{Stetcu_2011}). We therefore subtract the unkicked control trajectory frame by frame to isolate the response related exclusively to $\delta V_\sigma$, $\delta F_m(r,t) = F_m(r,t) - F_m^{(0)}(r,t)$. As a result, the excitation frequency spectrum $\omega(m)$ for $m>0$ can be extracted from the temporal Fourier transform of $\delta F_m$ over the $t\,\eF=965$ post-kick window. We retain only frequency peaks that display phase coherence across all ten radii, as a genuine collective displacement shifts the crystal uniformly across the ring width. The full details on the procedure are given in the End Matter.

As shown in Fig.~\ref{fig:goldstone}, the excitation dispersion is linear over $m=1\text{--}5$, where a weighted fit yields $\omega/\eF = 0.0198(11)\,m + 0.004(2)$. The fitted intercept lies within two standard errors of zero and is smaller than the $6.5\times10^{-3}\,\eF$ frequency resolution of the $t\,\eF=965$ record. Together with the sliding mode in Fig.~\ref{fig:order}(c), which shows that uniform translations cost zero energy, this indicates a gapless branch within our spectral resolution.

At $\Rmid$, the slope corresponds to a mode speed of $0.42\,v_F$ ($0.32$--$0.52\,v_F$ across the ring width), comparable to the first-sound speed of the uniform, unpolarized unitary gas at the same density, $c_1=\sqrt{\xi_B/3}\,v_F\approx0.35\,v_F$ (see End Matter), so the speed alone does not identify the mode, as supersolids support more than one sound branch~\cite{Sindik_2024, Hertkorn_2024}. Two observations tie the branch to the crystal. First, it is excited and detected in the spin channel, and the spin-density contrast is concentrated on the nodal lines, so $\delta F_m$ largely follows the displacement of the nodes. Second, because the spin-density crystal is periodic under $\theta\to\theta+2\pi/\nnod$, angular momentum is conserved modulo $\nnod$. An external drive at wavenumber $m$ consequently couples to the mode at $\vert{}m - \nnod\vert{}$ through angular Umklapp scattering, where the crystal lattice absorbs $\nnod$ units of angular momentum~\cite{Ashcroft_1976, Hertkorn_2024, Minakova2026}. In our simulations, pulses applied at $m=11$, $13$ and $14$ excite frequencies that agree with those of the $m=5$, $3$, and $2$ modes, respectively. This symmetry folding confirms a Brillouin-zone edge at $m = \nnod/2 = 8$, so the branch propagates on the crystalline lattice. The kick response weakens beyond $m=5$, with only weak coherent peaks near the extrapolated line at $m=6$ and $7$; at $m=8$ the pulse exerts no force on the nodes, which sit at extrema of the drive.

\ssec{Conclusions.}
Persistent currents induce spontaneous supersolid formation in a spin-imbalanced unitary Fermi gas and select the modulation period of the resulting crystal through quantized winding in addition to the polarization. Unlike most experimentally realized supersolids, where crystalline order arises from long-range dipolar interactions~\cite{TanziPRL_2019,Chomaz_2019,Bottcher_2019}, cavity-mediated coupling~\cite{Leonard_2017}, or engineered spin-orbit coupling~\cite{Li_2017}, here it emerges in a continuum Fermi gas with short-range interactions and is driven by a quantized persistent current. The period is quantized to an even number of nodes. The nodal lines act as $\pi$ junctions while the pairing field maintains global phase coherence around the ring. Once formed, the crystal slides freely, and its excitations form a linear branch consistent with a gapless mode, with the Brillouin-zone edge at $m=\nnod/2$ revealed by Umklapp folding. Quantized rotation, therefore, supplies an additional independent control parameter for inhomogeneous fermionic pairing; it actively organizes the crystalline state rather than merely probing it. Existing $^6$Li ring experiments provide phase-sensitive access to circulation~\cite{DelPace_2022, Pezze2024,xhaniPRA,Xhaniatoms}, which enables direct interference measurements of the nodal $\pi$ junctions and can furthermore resolve the density contrast between populations in two spin or hyperfine states. Finally, the Higgs mode~\cite{Behrle_2018} of the nodal structure and the dynamics of polarized vortices in the medium-polarization regime present primary targets for future investigation.

\ssec{Acknowledgments.}
We thank Alessio Recati and Elena Poli for fruitful discussions. This work was supported by the (Polish) National Science Center, under Grant Nos. UMO-2025/56/C/ST2/00420 (BT) and UMO-2022/45/B/ST2/00358 (GW). This publication was carried out with financial support from the Italian Ministry of Universities and Research (MUR) under the FIS 3 call D.D. No. 18010 of 12.11.2025. We acknowledge the Polish high-performance computing infrastructure PLGrid for awarding this project access to the LUMI supercomputer, owned by the EuroHPC Joint Undertaking and hosted by CSC (Finland) and the LUMI consortium, under PLL/2025/09/018958.

\bibliography{biblio}

\begin{thebibliography}{64}%
\makeatletter
\providecommand \@ifxundefined [1]{%
 \@ifx{#1\undefined}
}%
\providecommand \@ifnum [1]{%
 \ifnum #1\expandafter \@firstoftwo
 \else \expandafter \@secondoftwo
 \fi
}%
\providecommand \@ifx [1]{%
 \ifx #1\expandafter \@firstoftwo
 \else \expandafter \@secondoftwo
 \fi
}%
\providecommand \natexlab [1]{#1}%
\providecommand \enquote  [1]{``#1''}%
\providecommand \bibnamefont  [1]{#1}%
\providecommand \bibfnamefont [1]{#1}%
\providecommand \citenamefont [1]{#1}%
\providecommand \href@noop [0]{\@secondoftwo}%
\providecommand \href [0]{\begingroup \@sanitize@url \@href}%
\providecommand \@href[1]{\@@startlink{#1}\@@href}%
\providecommand \@@href[1]{\endgroup#1\@@endlink}%
\providecommand \@sanitize@url [0]{\catcode `\\12\catcode `\$12\catcode `\&12\catcode `\#12\catcode `\^12\catcode `\_12\catcode `\%12\relax}%
\providecommand \@@startlink[1]{}%
\providecommand \@@endlink[0]{}%
\providecommand \url  [0]{\begingroup\@sanitize@url \@url }%
\providecommand \@url [1]{\endgroup\@href {#1}{\urlprefix }}%
\providecommand \urlprefix  [0]{URL }%
\providecommand \Eprint [0]{\href }%
\providecommand \doibase [0]{http://dx.doi.org/}%
\providecommand \selectlanguage [0]{\@gobble}%
\providecommand \bibinfo  [0]{\@secondoftwo}%
\providecommand \bibfield  [0]{\@secondoftwo}%
\providecommand \translation [1]{[#1]}%
\providecommand \BibitemOpen [0]{}%
\providecommand \bibitemStop [0]{}%
\providecommand \bibitemNoStop [0]{.\EOS\space}%
\providecommand \EOS [0]{\spacefactor3000\relax}%
\providecommand \BibitemShut  [1]{\csname bibitem#1\endcsname}%
\let\auto@bib@innerbib\@empty
%</preamble>
\bibitem [{\citenamefont {Fulde}\ and\ \citenamefont {Ferrell}(1964)}]{Fulde_1964}%
  \BibitemOpen
  \bibfield  {author} {\bibinfo {author} {\bibfnamefont {Peter}\ \bibnamefont {Fulde}}\ and\ \bibinfo {author} {\bibfnamefont {Richard~A.}\ \bibnamefont {Ferrell}},\ }\bibfield  {title} {\enquote {\bibinfo {title} {{Superconductivity in a Strong Spin-Exchange Field}},}\ }\href {\doibase 10.1103/PhysRev.135.A550} {\bibfield  {journal} {\bibinfo  {journal} {Physical Review}\ }\textbf {\bibinfo {volume} {135}},\ \bibinfo {pages} {A550--A563} (\bibinfo {year} {1964})}\BibitemShut {NoStop}%
\bibitem [{\citenamefont {Larkin}\ and\ \citenamefont {Ovchinnikov}(1965)}]{Larkin_1965}%
  \BibitemOpen
  \bibfield  {author} {\bibinfo {author} {\bibfnamefont {A.~I.}\ \bibnamefont {Larkin}}\ and\ \bibinfo {author} {\bibfnamefont {Yu.~N.}\ \bibnamefont {Ovchinnikov}},\ }\bibfield  {title} {\enquote {\bibinfo {title} {{Nonuniform state of superconductors}},}\ }\href@noop {} {\bibfield  {journal} {\bibinfo  {journal} {Zh. Eksp. Teor. Fiz.}\ }\textbf {\bibinfo {volume} {47}},\ \bibinfo {pages} {1136} (\bibinfo {year} {1965})},\ \bibinfo {note} {[Sov. Phys. JETP {\bf 20}, 762 (1965)]}\BibitemShut {NoStop}%
\bibitem [{\citenamefont {Buzdin}(2005)}]{Buzdin_2005}%
  \BibitemOpen
  \bibfield  {author} {\bibinfo {author} {\bibfnamefont {A.~I.}\ \bibnamefont {Buzdin}},\ }\bibfield  {title} {\enquote {\bibinfo {title} {{Proximity effects in superconductor-ferromagnet heterostructures}},}\ }\href {\doibase 10.1103/RevModPhys.77.935} {\bibfield  {journal} {\bibinfo  {journal} {Reviews of Modern Physics}\ }\textbf {\bibinfo {volume} {77}},\ \bibinfo {pages} {935--976} (\bibinfo {year} {2005})}\BibitemShut {NoStop}%
\bibitem [{\citenamefont {Sedrakian}\ and\ \citenamefont {Clark}(2006)}]{Sedrakian_2006}%
  \BibitemOpen
  \bibfield  {author} {\bibinfo {author} {\bibfnamefont {Armen}\ \bibnamefont {Sedrakian}}\ and\ \bibinfo {author} {\bibfnamefont {John~W.}\ \bibnamefont {Clark}},\ }\bibfield  {title} {\enquote {\bibinfo {title} {{Nuclear Superconductivity in Compact Stars: BCS Theory and Beyond}},}\ }in\ \href {\doibase 10.1142/9789812773043_0006} {\emph {\bibinfo {booktitle} {Pairing in Fermionic Systems: Basic Concepts and Modern Applications}}},\ \bibinfo {series} {Series on Advances in Quantum Many-Body Theory}, Vol.~\bibinfo {volume} {8},\ \bibinfo {editor} {edited by\ \bibinfo {editor} {\bibfnamefont {Armen}\ \bibnamefont {Sedrakian}}, \bibinfo {editor} {\bibfnamefont {John~W.}\ \bibnamefont {Clark}}, \ and\ \bibinfo {editor} {\bibfnamefont {Mark}\ \bibnamefont {Alford}}}\ (\bibinfo  {publisher} {World Scientific},\ \bibinfo {address} {Singapore},\ \bibinfo {year} {2006})\ pp.\ \bibinfo {pages} {135--174}\BibitemShut {NoStop}%
\bibitem [{\citenamefont {Zwierlein}\ \emph {et~al.}(2006)\citenamefont {Zwierlein}, \citenamefont {Schirotzek}, \citenamefont {Schunck},\ and\ \citenamefont {Ketterle}}]{Zwierlein_2006}%
  \BibitemOpen
  \bibfield  {author} {\bibinfo {author} {\bibfnamefont {Martin~W.}\ \bibnamefont {Zwierlein}}, \bibinfo {author} {\bibfnamefont {Andre}\ \bibnamefont {Schirotzek}}, \bibinfo {author} {\bibfnamefont {Christian~H.}\ \bibnamefont {Schunck}}, \ and\ \bibinfo {author} {\bibfnamefont {Wolfgang}\ \bibnamefont {Ketterle}},\ }\bibfield  {title} {\enquote {\bibinfo {title} {{Fermionic superfluidity with imbalanced spin populations}},}\ }\href {\doibase 10.1126/science.1122318} {\bibfield  {journal} {\bibinfo  {journal} {Science}\ }\textbf {\bibinfo {volume} {311}},\ \bibinfo {pages} {492--496} (\bibinfo {year} {2006})}\BibitemShut {NoStop}%
\bibitem [{\citenamefont {Partridge}\ \emph {et~al.}(2006)\citenamefont {Partridge}, \citenamefont {Li}, \citenamefont {Kamar}, \citenamefont {Liao},\ and\ \citenamefont {Hulet}}]{Partridge_2006}%
  \BibitemOpen
  \bibfield  {author} {\bibinfo {author} {\bibfnamefont {G.~B.}\ \bibnamefont {Partridge}}, \bibinfo {author} {\bibfnamefont {W.}~\bibnamefont {Li}}, \bibinfo {author} {\bibfnamefont {R.~I.}\ \bibnamefont {Kamar}}, \bibinfo {author} {\bibfnamefont {Y.}~\bibnamefont {Liao}}, \ and\ \bibinfo {author} {\bibfnamefont {R.~G.}\ \bibnamefont {Hulet}},\ }\bibfield  {title} {\enquote {\bibinfo {title} {{Pairing and phase separation in a polarized Fermi gas}},}\ }\href {\doibase 10.1126/science.1122876} {\bibfield  {journal} {\bibinfo  {journal} {Science}\ }\textbf {\bibinfo {volume} {311}},\ \bibinfo {pages} {503--505} (\bibinfo {year} {2006})}\BibitemShut {NoStop}%
\bibitem [{\citenamefont {Stein}\ \emph {et~al.}(2014)\citenamefont {Stein}, \citenamefont {Sedrakian}, \citenamefont {Huang},\ and\ \citenamefont {Clark}}]{Stein_2014}%
  \BibitemOpen
  \bibfield  {author} {\bibinfo {author} {\bibfnamefont {Martin}\ \bibnamefont {Stein}}, \bibinfo {author} {\bibfnamefont {Armen}\ \bibnamefont {Sedrakian}}, \bibinfo {author} {\bibfnamefont {Xu-Guang}\ \bibnamefont {Huang}}, \ and\ \bibinfo {author} {\bibfnamefont {John~W.}\ \bibnamefont {Clark}},\ }\bibfield  {title} {\enquote {\bibinfo {title} {{BCS-BEC crossovers and unconventional phases in dilute nuclear matter}},}\ }\href {\doibase 10.1103/physrevc.90.065804} {\bibfield  {journal} {\bibinfo  {journal} {Physical Review C}\ }\textbf {\bibinfo {volume} {90}},\ \bibinfo {pages} {065804} (\bibinfo {year} {2014})}\BibitemShut {NoStop}%
\bibitem [{\citenamefont {Kinnunen}\ \emph {et~al.}(2018)\citenamefont {Kinnunen}, \citenamefont {Baarsma}, \citenamefont {Martikainen},\ and\ \citenamefont {T\"orm\"a}}]{Torma_2018}%
  \BibitemOpen
  \bibfield  {author} {\bibinfo {author} {\bibfnamefont {Jami~J.}\ \bibnamefont {Kinnunen}}, \bibinfo {author} {\bibfnamefont {Jildou~E.}\ \bibnamefont {Baarsma}}, \bibinfo {author} {\bibfnamefont {Jani-Petri}\ \bibnamefont {Martikainen}}, \ and\ \bibinfo {author} {\bibfnamefont {P\"aivi}\ \bibnamefont {T\"orm\"a}},\ }\bibfield  {title} {\enquote {\bibinfo {title} {{The Fulde--Ferrell--Larkin--Ovchinnikov state for ultracold fermions in lattice and harmonic potentials: a review}},}\ }\href {\doibase 10.1088/1361-6633/aaa4ad} {\bibfield  {journal} {\bibinfo  {journal} {Reports on Progress in Physics}\ }\textbf {\bibinfo {volume} {81}},\ \bibinfo {pages} {046401} (\bibinfo {year} {2018})}\BibitemShut {NoStop}%
\bibitem [{\citenamefont {Pirolo}\ \emph {et~al.}(2026)\citenamefont {Pirolo}, \citenamefont {Pisani},\ and\ \citenamefont {Pieri}}]{Pirolo_2026}%
  \BibitemOpen
  \bibfield  {author} {\bibinfo {author} {\bibfnamefont {Francesco}\ \bibnamefont {Pirolo}}, \bibinfo {author} {\bibfnamefont {Leonardo}\ \bibnamefont {Pisani}}, \ and\ \bibinfo {author} {\bibfnamefont {Pierbiagio}\ \bibnamefont {Pieri}},\ }\bibfield  {title} {\enquote {\bibinfo {title} {{Fulde-Ferrell-Larkin-Ovchinnikov transition and quantum criticality in polarized Fermi gases}},}\ }\href {\doibase 10.1103/4wwz-xg5q} {\bibfield  {journal} {\bibinfo  {journal} {Phys. Rev. B}\ }\textbf {\bibinfo {volume} {114}},\ \bibinfo {pages} {254507} (\bibinfo {year} {2026})}\BibitemShut {NoStop}%
\bibitem [{\citenamefont {Mayaffre}\ \emph {et~al.}(2014)\citenamefont {Mayaffre}, \citenamefont {Kr{\"a}mer}, \citenamefont {Horvati{\'c}}, \citenamefont {Berthier}, \citenamefont {Miyagawa}, \citenamefont {Kanoda},\ and\ \citenamefont {Mitrovi{\'c}}}]{Mayaffre_2014}%
  \BibitemOpen
  \bibfield  {author} {\bibinfo {author} {\bibfnamefont {H.}~\bibnamefont {Mayaffre}}, \bibinfo {author} {\bibfnamefont {S.}~\bibnamefont {Kr{\"a}mer}}, \bibinfo {author} {\bibfnamefont {M.}~\bibnamefont {Horvati{\'c}}}, \bibinfo {author} {\bibfnamefont {C.}~\bibnamefont {Berthier}}, \bibinfo {author} {\bibfnamefont {K.}~\bibnamefont {Miyagawa}}, \bibinfo {author} {\bibfnamefont {K.}~\bibnamefont {Kanoda}}, \ and\ \bibinfo {author} {\bibfnamefont {V.~F.}\ \bibnamefont {Mitrovi{\'c}}},\ }\bibfield  {title} {\enquote {\bibinfo {title} {{Evidence of Andreev bound states as a hallmark of the FFLO phase in $\kappa$-(BEDT-TTF)$_2$Cu(NCS)$_2$}},}\ }\href {\doibase 10.1038/nphys3121} {\bibfield  {journal} {\bibinfo  {journal} {Nature Physics}\ }\textbf {\bibinfo {volume} {10}},\ \bibinfo {pages} {928--932} (\bibinfo {year} {2014})}\BibitemShut {NoStop}%
\bibitem [{\citenamefont {Liao}\ \emph {et~al.}(2010)\citenamefont {Liao}, \citenamefont {Rittner}, \citenamefont {Paprotta}, \citenamefont {Li}, \citenamefont {Partridge}, \citenamefont {Hulet}, \citenamefont {Baur},\ and\ \citenamefont {Mueller}}]{Liao_2010}%
  \BibitemOpen
  \bibfield  {author} {\bibinfo {author} {\bibfnamefont {{Yean-an}}\ \bibnamefont {Liao}}, \bibinfo {author} {\bibfnamefont {Ann Sophie~C.}\ \bibnamefont {Rittner}}, \bibinfo {author} {\bibfnamefont {Tobias}\ \bibnamefont {Paprotta}}, \bibinfo {author} {\bibfnamefont {Wenhui}\ \bibnamefont {Li}}, \bibinfo {author} {\bibfnamefont {Guthrie~B.}\ \bibnamefont {Partridge}}, \bibinfo {author} {\bibfnamefont {Randall~G.}\ \bibnamefont {Hulet}}, \bibinfo {author} {\bibfnamefont {Stefan~K.}\ \bibnamefont {Baur}}, \ and\ \bibinfo {author} {\bibfnamefont {Erich~J.}\ \bibnamefont {Mueller}},\ }\bibfield  {title} {\enquote {\bibinfo {title} {{Spin-imbalance in a one-dimensional Fermi gas}},}\ }\href {\doibase 10.1038/nature09393} {\bibfield  {journal} {\bibinfo  {journal} {Nature}\ }\textbf {\bibinfo {volume} {467}},\ \bibinfo {pages} {567--569} (\bibinfo {year} {2010})}\BibitemShut {NoStop}%
\bibitem [{\citenamefont {Radzihovsky}\ and\ \citenamefont {Vishwanath}(2009)}]{Radzihovsky_2009}%
  \BibitemOpen
  \bibfield  {author} {\bibinfo {author} {\bibfnamefont {Leo}\ \bibnamefont {Radzihovsky}}\ and\ \bibinfo {author} {\bibfnamefont {Ashvin}\ \bibnamefont {Vishwanath}},\ }\bibfield  {title} {\enquote {\bibinfo {title} {{Quantum Liquid Crystals in an Imbalanced Fermi Gas: Fluctuations and Fractional Vortices in Larkin-Ovchinnikov States}},}\ }\href {\doibase 10.1103/PhysRevLett.103.010404} {\bibfield  {journal} {\bibinfo  {journal} {Physical Review Letters}\ }\textbf {\bibinfo {volume} {103}},\ \bibinfo {pages} {010404} (\bibinfo {year} {2009})}\BibitemShut {NoStop}%
\bibitem [{\citenamefont {Radzihovsky}\ and\ \citenamefont {Sheehy}(2010)}]{Radzihovsky_2010}%
  \BibitemOpen
  \bibfield  {author} {\bibinfo {author} {\bibfnamefont {Leo}\ \bibnamefont {Radzihovsky}}\ and\ \bibinfo {author} {\bibfnamefont {Daniel~E.}\ \bibnamefont {Sheehy}},\ }\bibfield  {title} {\enquote {\bibinfo {title} {{Imbalanced Feshbach-resonant Fermi gases}},}\ }\href {\doibase 10.1088/0034-4885/73/7/076501} {\bibfield  {journal} {\bibinfo  {journal} {Reports on Progress in Physics}\ }\textbf {\bibinfo {volume} {73}},\ \bibinfo {pages} {076501} (\bibinfo {year} {2010})}\BibitemShut {NoStop}%
\bibitem [{\citenamefont {Radzihovsky}(2011)}]{Radzihovsky_2011}%
  \BibitemOpen
  \bibfield  {author} {\bibinfo {author} {\bibfnamefont {Leo}\ \bibnamefont {Radzihovsky}},\ }\bibfield  {title} {\enquote {\bibinfo {title} {{Fluctuations and phase transitions in Larkin-Ovchinnikov liquid-crystal states of a population-imbalanced resonant Fermi gas}},}\ }\href {\doibase 10.1103/PhysRevA.84.023611} {\bibfield  {journal} {\bibinfo  {journal} {Physical Review A}\ }\textbf {\bibinfo {volume} {84}},\ \bibinfo {pages} {023611} (\bibinfo {year} {2011})}\BibitemShut {NoStop}%
\bibitem [{\citenamefont {Bulgac}\ and\ \citenamefont {Forbes}(2008)}]{Bulgac_2008}%
  \BibitemOpen
  \bibfield  {author} {\bibinfo {author} {\bibfnamefont {Aurel}\ \bibnamefont {Bulgac}}\ and\ \bibinfo {author} {\bibfnamefont {Michael~McNeil}\ \bibnamefont {Forbes}},\ }\bibfield  {title} {\enquote {\bibinfo {title} {{Unitary Fermi Supersolid: The Larkin-Ovchinnikov Phase}},}\ }\href {\doibase 10.1103/PhysRevLett.101.215301} {\bibfield  {journal} {\bibinfo  {journal} {Physical Review Letters}\ }\textbf {\bibinfo {volume} {101}},\ \bibinfo {pages} {215301} (\bibinfo {year} {2008})}\BibitemShut {NoStop}%
\bibitem [{\citenamefont {Magierski}\ \emph {et~al.}(2019)\citenamefont {Magierski}, \citenamefont {T\"uzemen},\ and\ \citenamefont {Wlaz\l{}owski}}]{Magierski_2019}%
  \BibitemOpen
  \bibfield  {author} {\bibinfo {author} {\bibfnamefont {Piotr}\ \bibnamefont {Magierski}}, \bibinfo {author} {\bibfnamefont {Bu\u{g}ra}\ \bibnamefont {T\"uzemen}}, \ and\ \bibinfo {author} {\bibfnamefont {Gabriel}\ \bibnamefont {Wlaz\l{}owski}},\ }\bibfield  {title} {\enquote {\bibinfo {title} {{Spin-polarized droplets in the unitary Fermi gas}},}\ }\href {\doibase 10.1103/PhysRevA.100.033613} {\bibfield  {journal} {\bibinfo  {journal} {Physical Review A}\ }\textbf {\bibinfo {volume} {100}},\ \bibinfo {pages} {033613} (\bibinfo {year} {2019})}\BibitemShut {NoStop}%
\bibitem [{\citenamefont {T\"uzemen}\ \emph {et~al.}(2023)\citenamefont {T\"uzemen}, \citenamefont {Zawi\'slak}, \citenamefont {Wlaz\l{}owski},\ and\ \citenamefont {Magierski}}]{Tuzemen_2023}%
  \BibitemOpen
  \bibfield  {author} {\bibinfo {author} {\bibfnamefont {Bu\u{g}ra}\ \bibnamefont {T\"uzemen}}, \bibinfo {author} {\bibfnamefont {Tomasz}\ \bibnamefont {Zawi\'slak}}, \bibinfo {author} {\bibfnamefont {Gabriel}\ \bibnamefont {Wlaz\l{}owski}}, \ and\ \bibinfo {author} {\bibfnamefont {Piotr}\ \bibnamefont {Magierski}},\ }\bibfield  {title} {\enquote {\bibinfo {title} {{Disordered structures in ultracold spin-imbalanced Fermi gas}},}\ }\href {\doibase 10.1088/1367-2630/acc26b} {\bibfield  {journal} {\bibinfo  {journal} {New Journal of Physics}\ }\textbf {\bibinfo {volume} {25}},\ \bibinfo {pages} {033013} (\bibinfo {year} {2023})}\BibitemShut {NoStop}%
\bibitem [{\citenamefont {L{\'e}onard}\ \emph {et~al.}(2017{\natexlab{a}})\citenamefont {L{\'e}onard}, \citenamefont {Morales}, \citenamefont {Zupancic}, \citenamefont {Esslinger},\ and\ \citenamefont {Donner}}]{Leonard_2017}%
  \BibitemOpen
  \bibfield  {author} {\bibinfo {author} {\bibfnamefont {Julian}\ \bibnamefont {L{\'e}onard}}, \bibinfo {author} {\bibfnamefont {Andrea}\ \bibnamefont {Morales}}, \bibinfo {author} {\bibfnamefont {Philip}\ \bibnamefont {Zupancic}}, \bibinfo {author} {\bibfnamefont {Tilman}\ \bibnamefont {Esslinger}}, \ and\ \bibinfo {author} {\bibfnamefont {Tobias}\ \bibnamefont {Donner}},\ }\bibfield  {title} {\enquote {\bibinfo {title} {{Supersolid formation in a quantum gas breaking a continuous translational symmetry}},}\ }\href {\doibase 10.1038/nature21067} {\bibfield  {journal} {\bibinfo  {journal} {Nature}\ }\textbf {\bibinfo {volume} {543}},\ \bibinfo {pages} {87--90} (\bibinfo {year} {2017}{\natexlab{a}})}\BibitemShut {NoStop}%
\bibitem [{\citenamefont {Li}\ \emph {et~al.}(2017)\citenamefont {Li}, \citenamefont {Lee}, \citenamefont {Huang}, \citenamefont {Burchesky}, \citenamefont {Shteynas}, \citenamefont {Top}, \citenamefont {Jamison},\ and\ \citenamefont {Ketterle}}]{Li_2017}%
  \BibitemOpen
  \bibfield  {author} {\bibinfo {author} {\bibfnamefont {Jun-Ru}\ \bibnamefont {Li}}, \bibinfo {author} {\bibfnamefont {Jeongwon}\ \bibnamefont {Lee}}, \bibinfo {author} {\bibfnamefont {Wujie}\ \bibnamefont {Huang}}, \bibinfo {author} {\bibfnamefont {Sean}\ \bibnamefont {Burchesky}}, \bibinfo {author} {\bibfnamefont {Boris}\ \bibnamefont {Shteynas}}, \bibinfo {author} {\bibfnamefont {Furkan~{\c{C}}a{\u{g}}r{\i}}\ \bibnamefont {Top}}, \bibinfo {author} {\bibfnamefont {Alan~O.}\ \bibnamefont {Jamison}}, \ and\ \bibinfo {author} {\bibfnamefont {Wolfgang}\ \bibnamefont {Ketterle}},\ }\bibfield  {title} {\enquote {\bibinfo {title} {{A stripe phase with supersolid properties in spin--orbit-coupled Bose--Einstein condensates}},}\ }\href {\doibase 10.1038/nature21431} {\bibfield  {journal} {\bibinfo  {journal} {Nature}\ }\textbf {\bibinfo {volume} {543}},\ \bibinfo {pages} {91--94} (\bibinfo {year} {2017})}\BibitemShut {NoStop}%
\bibitem [{\citenamefont {B{\"o}ttcher}\ \emph {et~al.}(2019)\citenamefont {B{\"o}ttcher}, \citenamefont {Schmidt}, \citenamefont {Wenzel}, \citenamefont {Hertkorn}, \citenamefont {Guo}, \citenamefont {Langen},\ and\ \citenamefont {Pfau}}]{Bottcher_2019}%
  \BibitemOpen
  \bibfield  {author} {\bibinfo {author} {\bibfnamefont {Fabian}\ \bibnamefont {B{\"o}ttcher}}, \bibinfo {author} {\bibfnamefont {Jan-Niklas}\ \bibnamefont {Schmidt}}, \bibinfo {author} {\bibfnamefont {Matthias}\ \bibnamefont {Wenzel}}, \bibinfo {author} {\bibfnamefont {Jens}\ \bibnamefont {Hertkorn}}, \bibinfo {author} {\bibfnamefont {Mingyang}\ \bibnamefont {Guo}}, \bibinfo {author} {\bibfnamefont {Tim}\ \bibnamefont {Langen}}, \ and\ \bibinfo {author} {\bibfnamefont {Tilman}\ \bibnamefont {Pfau}},\ }\bibfield  {title} {\enquote {\bibinfo {title} {{Transient Supersolid Properties in an Array of Dipolar Quantum Droplets}},}\ }\href {\doibase 10.1103/physrevx.9.011051} {\bibfield  {journal} {\bibinfo  {journal} {Physical Review X}\ }\textbf {\bibinfo {volume} {9}},\ \bibinfo {pages} {011051} (\bibinfo {year} {2019})}\BibitemShut {NoStop}%
\bibitem [{\citenamefont {Chomaz}\ \emph {et~al.}(2019)\citenamefont {Chomaz}, \citenamefont {Petter}, \citenamefont {Ilzh{\"o}fer}, \citenamefont {Natale}, \citenamefont {Trautmann}, \citenamefont {Politi}, \citenamefont {Durastante}, \citenamefont {van Bijnen}, \citenamefont {Patscheider}, \citenamefont {Sohmen}, \citenamefont {Mark},\ and\ \citenamefont {Ferlaino}}]{Chomaz_2019}%
  \BibitemOpen
  \bibfield  {author} {\bibinfo {author} {\bibfnamefont {L.}~\bibnamefont {Chomaz}}, \bibinfo {author} {\bibfnamefont {D.}~\bibnamefont {Petter}}, \bibinfo {author} {\bibfnamefont {P.}~\bibnamefont {Ilzh{\"o}fer}}, \bibinfo {author} {\bibfnamefont {G.}~\bibnamefont {Natale}}, \bibinfo {author} {\bibfnamefont {A.}~\bibnamefont {Trautmann}}, \bibinfo {author} {\bibfnamefont {C.}~\bibnamefont {Politi}}, \bibinfo {author} {\bibfnamefont {G.}~\bibnamefont {Durastante}}, \bibinfo {author} {\bibfnamefont {R.M.W.}\ \bibnamefont {van Bijnen}}, \bibinfo {author} {\bibfnamefont {A.}~\bibnamefont {Patscheider}}, \bibinfo {author} {\bibfnamefont {M.}~\bibnamefont {Sohmen}}, \bibinfo {author} {\bibfnamefont {M.J.}\ \bibnamefont {Mark}}, \ and\ \bibinfo {author} {\bibfnamefont {F.}~\bibnamefont {Ferlaino}},\ }\bibfield  {title} {\enquote {\bibinfo {title} {{Long-Lived and Transient Supersolid Behaviors in Dipolar Quantum Gases}},}\ }\href {\doibase 10.1103/physrevx.9.021012} {\bibfield  {journal} {\bibinfo  {journal}
  {Physical Review X}\ }\textbf {\bibinfo {volume} {9}},\ \bibinfo {pages} {021012} (\bibinfo {year} {2019})}\BibitemShut {NoStop}%
\bibitem [{\citenamefont {Tanzi}\ \emph {et~al.}(2019{\natexlab{a}})\citenamefont {Tanzi}, \citenamefont {Lucioni}, \citenamefont {Fam{\`a}}, \citenamefont {Catani}, \citenamefont {Fioretti}, \citenamefont {Gabbanini}, \citenamefont {Bisset}, \citenamefont {Santos},\ and\ \citenamefont {Modugno}}]{TanziPRL_2019}%
  \BibitemOpen
  \bibfield  {author} {\bibinfo {author} {\bibfnamefont {L.}~\bibnamefont {Tanzi}}, \bibinfo {author} {\bibfnamefont {E.}~\bibnamefont {Lucioni}}, \bibinfo {author} {\bibfnamefont {F.}~\bibnamefont {Fam{\`a}}}, \bibinfo {author} {\bibfnamefont {J.}~\bibnamefont {Catani}}, \bibinfo {author} {\bibfnamefont {A.}~\bibnamefont {Fioretti}}, \bibinfo {author} {\bibfnamefont {C.}~\bibnamefont {Gabbanini}}, \bibinfo {author} {\bibfnamefont {R.N.}\ \bibnamefont {Bisset}}, \bibinfo {author} {\bibfnamefont {L.}~\bibnamefont {Santos}}, \ and\ \bibinfo {author} {\bibfnamefont {G.}~\bibnamefont {Modugno}},\ }\bibfield  {title} {\enquote {\bibinfo {title} {{Observation of a Dipolar Quantum Gas with Metastable Supersolid Properties}},}\ }\href {\doibase 10.1103/physrevlett.122.130405} {\bibfield  {journal} {\bibinfo  {journal} {Physical Review Letters}\ }\textbf {\bibinfo {volume} {122}},\ \bibinfo {pages} {130405} (\bibinfo {year} {2019}{\natexlab{a}})}\BibitemShut {NoStop}%
\bibitem [{\citenamefont {Norcia}\ \emph {et~al.}(2021)\citenamefont {Norcia}, \citenamefont {Politi}, \citenamefont {Klaus}, \citenamefont {Poli}, \citenamefont {Sohmen}, \citenamefont {Mark}, \citenamefont {Bisset}, \citenamefont {Santos},\ and\ \citenamefont {Ferlaino}}]{Norcia_2021}%
  \BibitemOpen
  \bibfield  {author} {\bibinfo {author} {\bibfnamefont {Matthew~A.}\ \bibnamefont {Norcia}}, \bibinfo {author} {\bibfnamefont {Claudia}\ \bibnamefont {Politi}}, \bibinfo {author} {\bibfnamefont {Lauritz}\ \bibnamefont {Klaus}}, \bibinfo {author} {\bibfnamefont {Elena}\ \bibnamefont {Poli}}, \bibinfo {author} {\bibfnamefont {Maximilian}\ \bibnamefont {Sohmen}}, \bibinfo {author} {\bibfnamefont {Manfred~J.}\ \bibnamefont {Mark}}, \bibinfo {author} {\bibfnamefont {Russell~N.}\ \bibnamefont {Bisset}}, \bibinfo {author} {\bibfnamefont {Luis}\ \bibnamefont {Santos}}, \ and\ \bibinfo {author} {\bibfnamefont {Francesca}\ \bibnamefont {Ferlaino}},\ }\bibfield  {title} {\enquote {\bibinfo {title} {{Two-dimensional supersolidity in a dipolar quantum gas}},}\ }\href {\doibase 10.1038/s41586-021-03725-7} {\bibfield  {journal} {\bibinfo  {journal} {Nature}\ }\textbf {\bibinfo {volume} {596}},\ \bibinfo {pages} {357--361} (\bibinfo {year} {2021})}\BibitemShut {NoStop}%
\bibitem [{\citenamefont {Recati}\ and\ \citenamefont {Stringari}(2023)}]{Recati_2023}%
  \BibitemOpen
  \bibfield  {author} {\bibinfo {author} {\bibfnamefont {Alessio}\ \bibnamefont {Recati}}\ and\ \bibinfo {author} {\bibfnamefont {Sandro}\ \bibnamefont {Stringari}},\ }\bibfield  {title} {\enquote {\bibinfo {title} {{Supersolidity in ultracold dipolar gases}},}\ }\href {\doibase 10.1038/s42254-023-00648-2} {\bibfield  {journal} {\bibinfo  {journal} {Nature Reviews Physics}\ }\textbf {\bibinfo {volume} {5}},\ \bibinfo {pages} {735--743} (\bibinfo {year} {2023})}\BibitemShut {NoStop}%
\bibitem [{\citenamefont {Guo}\ \emph {et~al.}(2019)\citenamefont {Guo}, \citenamefont {B\"ottcher}, \citenamefont {Hertkorn}, \citenamefont {Schmidt}, \citenamefont {Wenzel}, \citenamefont {B\"uchler}, \citenamefont {Langen},\ and\ \citenamefont {Pfau}}]{Guo_2019}%
  \BibitemOpen
  \bibfield  {author} {\bibinfo {author} {\bibfnamefont {Mingyang}\ \bibnamefont {Guo}}, \bibinfo {author} {\bibfnamefont {Fabian}\ \bibnamefont {B\"ottcher}}, \bibinfo {author} {\bibfnamefont {Jens}\ \bibnamefont {Hertkorn}}, \bibinfo {author} {\bibfnamefont {Jan-Niklas}\ \bibnamefont {Schmidt}}, \bibinfo {author} {\bibfnamefont {Matthias}\ \bibnamefont {Wenzel}}, \bibinfo {author} {\bibfnamefont {Hans~Peter}\ \bibnamefont {B\"uchler}}, \bibinfo {author} {\bibfnamefont {Tim}\ \bibnamefont {Langen}}, \ and\ \bibinfo {author} {\bibfnamefont {Tilman}\ \bibnamefont {Pfau}},\ }\bibfield  {title} {\enquote {\bibinfo {title} {{The low-energy Goldstone mode in a trapped dipolar supersolid}},}\ }\href {\doibase 10.1038/s41586-019-1569-5} {\bibfield  {journal} {\bibinfo  {journal} {Nature}\ }\textbf {\bibinfo {volume} {574}},\ \bibinfo {pages} {386--389} (\bibinfo {year} {2019})}\BibitemShut {NoStop}%
\bibitem [{\citenamefont {Tanzi}\ \emph {et~al.}(2019{\natexlab{b}})\citenamefont {Tanzi}, \citenamefont {Roccuzzo}, \citenamefont {Lucioni}, \citenamefont {Fam\`a}, \citenamefont {Fioretti}, \citenamefont {Gabbanini}, \citenamefont {Modugno}, \citenamefont {Recati},\ and\ \citenamefont {Stringari}}]{Tanzi_2019}%
  \BibitemOpen
  \bibfield  {author} {\bibinfo {author} {\bibfnamefont {L.}~\bibnamefont {Tanzi}}, \bibinfo {author} {\bibfnamefont {S.~M.}\ \bibnamefont {Roccuzzo}}, \bibinfo {author} {\bibfnamefont {E.}~\bibnamefont {Lucioni}}, \bibinfo {author} {\bibfnamefont {F.}~\bibnamefont {Fam\`a}}, \bibinfo {author} {\bibfnamefont {A.}~\bibnamefont {Fioretti}}, \bibinfo {author} {\bibfnamefont {C.}~\bibnamefont {Gabbanini}}, \bibinfo {author} {\bibfnamefont {G.}~\bibnamefont {Modugno}}, \bibinfo {author} {\bibfnamefont {A.}~\bibnamefont {Recati}}, \ and\ \bibinfo {author} {\bibfnamefont {S.}~\bibnamefont {Stringari}},\ }\bibfield  {title} {\enquote {\bibinfo {title} {{Supersolid symmetry breaking from compressional oscillations in a dipolar quantum gas}},}\ }\href {\doibase 10.1038/s41586-019-1568-6} {\bibfield  {journal} {\bibinfo  {journal} {Nature}\ }\textbf {\bibinfo {volume} {574}},\ \bibinfo {pages} {382--385} (\bibinfo {year} {2019}{\natexlab{b}})}\BibitemShut {NoStop}%
\bibitem [{\citenamefont {Natale}\ \emph {et~al.}(2019)\citenamefont {Natale}, \citenamefont {van Bijnen}, \citenamefont {Patscheider}, \citenamefont {Petter}, \citenamefont {Mark}, \citenamefont {Chomaz},\ and\ \citenamefont {Ferlaino}}]{Natale_2019}%
  \BibitemOpen
  \bibfield  {author} {\bibinfo {author} {\bibfnamefont {G.}~\bibnamefont {Natale}}, \bibinfo {author} {\bibfnamefont {R.M.W.}\ \bibnamefont {van Bijnen}}, \bibinfo {author} {\bibfnamefont {A.}~\bibnamefont {Patscheider}}, \bibinfo {author} {\bibfnamefont {D.}~\bibnamefont {Petter}}, \bibinfo {author} {\bibfnamefont {M.J.}\ \bibnamefont {Mark}}, \bibinfo {author} {\bibfnamefont {L.}~\bibnamefont {Chomaz}}, \ and\ \bibinfo {author} {\bibfnamefont {F.}~\bibnamefont {Ferlaino}},\ }\bibfield  {title} {\enquote {\bibinfo {title} {{Excitation Spectrum of a Trapped Dipolar Supersolid and Its Experimental Evidence}},}\ }\href {\doibase 10.1103/physrevlett.123.050402} {\bibfield  {journal} {\bibinfo  {journal} {Physical Review Letters}\ }\textbf {\bibinfo {volume} {123}},\ \bibinfo {pages} {050402} (\bibinfo {year} {2019})}\BibitemShut {NoStop}%
\bibitem [{\citenamefont {L{\'e}onard}\ \emph {et~al.}(2017{\natexlab{b}})\citenamefont {L{\'e}onard}, \citenamefont {Morales}, \citenamefont {Zupancic}, \citenamefont {Donner},\ and\ \citenamefont {Esslinger}}]{Leonard_2017b}%
  \BibitemOpen
  \bibfield  {author} {\bibinfo {author} {\bibfnamefont {Julian}\ \bibnamefont {L{\'e}onard}}, \bibinfo {author} {\bibfnamefont {Andrea}\ \bibnamefont {Morales}}, \bibinfo {author} {\bibfnamefont {Philip}\ \bibnamefont {Zupancic}}, \bibinfo {author} {\bibfnamefont {Tobias}\ \bibnamefont {Donner}}, \ and\ \bibinfo {author} {\bibfnamefont {Tilman}\ \bibnamefont {Esslinger}},\ }\bibfield  {title} {\enquote {\bibinfo {title} {{Monitoring and manipulating Higgs and Goldstone modes in a supersolid quantum gas}},}\ }\href {\doibase 10.1126/science.aan2608} {\bibfield  {journal} {\bibinfo  {journal} {Science}\ }\textbf {\bibinfo {volume} {358}},\ \bibinfo {pages} {1415--1418} (\bibinfo {year} {2017}{\natexlab{b}})}\BibitemShut {NoStop}%
\bibitem [{\citenamefont {Hertkorn}\ \emph {et~al.}(2021)\citenamefont {Hertkorn}, \citenamefont {Schmidt}, \citenamefont {B{\"o}ttcher}, \citenamefont {Guo}, \citenamefont {Schmidt}, \citenamefont {Ng}, \citenamefont {Graham}, \citenamefont {B{\"u}chler}, \citenamefont {Langen}, \citenamefont {Zwierlein},\ and\ \citenamefont {Pfau}}]{Hertkorn_2021}%
  \BibitemOpen
  \bibfield  {author} {\bibinfo {author} {\bibfnamefont {J.}~\bibnamefont {Hertkorn}}, \bibinfo {author} {\bibfnamefont {J.-N.}\ \bibnamefont {Schmidt}}, \bibinfo {author} {\bibfnamefont {F.}~\bibnamefont {B{\"o}ttcher}}, \bibinfo {author} {\bibfnamefont {M.}~\bibnamefont {Guo}}, \bibinfo {author} {\bibfnamefont {M.}~\bibnamefont {Schmidt}}, \bibinfo {author} {\bibfnamefont {K.S.H.}\ \bibnamefont {Ng}}, \bibinfo {author} {\bibfnamefont {S.D.}\ \bibnamefont {Graham}}, \bibinfo {author} {\bibfnamefont {H.P.}\ \bibnamefont {B{\"u}chler}}, \bibinfo {author} {\bibfnamefont {T.}~\bibnamefont {Langen}}, \bibinfo {author} {\bibfnamefont {M.}~\bibnamefont {Zwierlein}}, \ and\ \bibinfo {author} {\bibfnamefont {T.}~\bibnamefont {Pfau}},\ }\bibfield  {title} {\enquote {\bibinfo {title} {{Density Fluctuations across the Superfluid-Supersolid Phase Transition in a Dipolar Quantum Gas}},}\ }\href {\doibase 10.1103/physrevx.11.011037} {\bibfield  {journal} {\bibinfo  {journal} {Physical Review X}\ }\textbf {\bibinfo {volume}
  {11}},\ \bibinfo {pages} {011037} (\bibinfo {year} {2021})}\BibitemShut {NoStop}%
\bibitem [{\citenamefont {Geier}\ \emph {et~al.}(2021)\citenamefont {Geier}, \citenamefont {Martone}, \citenamefont {Hauke},\ and\ \citenamefont {Stringari}}]{Geier_2021}%
  \BibitemOpen
  \bibfield  {author} {\bibinfo {author} {\bibfnamefont {Kevin~T.}\ \bibnamefont {Geier}}, \bibinfo {author} {\bibfnamefont {Giovanni~I.}\ \bibnamefont {Martone}}, \bibinfo {author} {\bibfnamefont {Philipp}\ \bibnamefont {Hauke}}, \ and\ \bibinfo {author} {\bibfnamefont {Sandro}\ \bibnamefont {Stringari}},\ }\bibfield  {title} {\enquote {\bibinfo {title} {{Exciting the Goldstone Modes of a Supersolid Spin-Orbit-Coupled Bose Gas}},}\ }\href {\doibase 10.1103/PhysRevLett.127.115301} {\bibfield  {journal} {\bibinfo  {journal} {Physical Review Letters}\ }\textbf {\bibinfo {volume} {127}},\ \bibinfo {pages} {115301} (\bibinfo {year} {2021})}\BibitemShut {NoStop}%
\bibitem [{\citenamefont {Biagioni}\ \emph {et~al.}(2024)\citenamefont {Biagioni}, \citenamefont {Antolini}, \citenamefont {Donelli}, \citenamefont {Pezz{\`e}}, \citenamefont {Smerzi}, \citenamefont {Fattori}, \citenamefont {Fioretti}, \citenamefont {Gabbanini}, \citenamefont {Inguscio}, \citenamefont {Tanzi},\ and\ \citenamefont {Modugno}}]{Biagioni_2024}%
  \BibitemOpen
  \bibfield  {author} {\bibinfo {author} {\bibfnamefont {G.}~\bibnamefont {Biagioni}}, \bibinfo {author} {\bibfnamefont {N.}~\bibnamefont {Antolini}}, \bibinfo {author} {\bibfnamefont {B.}~\bibnamefont {Donelli}}, \bibinfo {author} {\bibfnamefont {L.}~\bibnamefont {Pezz{\`e}}}, \bibinfo {author} {\bibfnamefont {A.}~\bibnamefont {Smerzi}}, \bibinfo {author} {\bibfnamefont {M.}~\bibnamefont {Fattori}}, \bibinfo {author} {\bibfnamefont {A.}~\bibnamefont {Fioretti}}, \bibinfo {author} {\bibfnamefont {C.}~\bibnamefont {Gabbanini}}, \bibinfo {author} {\bibfnamefont {M.}~\bibnamefont {Inguscio}}, \bibinfo {author} {\bibfnamefont {L.}~\bibnamefont {Tanzi}}, \ and\ \bibinfo {author} {\bibfnamefont {G.}~\bibnamefont {Modugno}},\ }\bibfield  {title} {\enquote {\bibinfo {title} {{Measurement of the superfluid fraction of a supersolid by Josephson effect}},}\ }\href {\doibase 10.1038/s41586-024-07361-9} {\bibfield  {journal} {\bibinfo  {journal} {Nature}\ }\textbf {\bibinfo {volume} {629}},\ \bibinfo {pages} {773--777}
  (\bibinfo {year} {2024})}\BibitemShut {NoStop}%
\bibitem [{\citenamefont {Tanzi}\ \emph {et~al.}(2021)\citenamefont {Tanzi}, \citenamefont {Maloberti}, \citenamefont {Biagioni}, \citenamefont {Fioretti}, \citenamefont {Gabbanini},\ and\ \citenamefont {Modugno}}]{Tanzi_2021}%
  \BibitemOpen
  \bibfield  {author} {\bibinfo {author} {\bibfnamefont {L.}~\bibnamefont {Tanzi}}, \bibinfo {author} {\bibfnamefont {J.~G.}\ \bibnamefont {Maloberti}}, \bibinfo {author} {\bibfnamefont {G.}~\bibnamefont {Biagioni}}, \bibinfo {author} {\bibfnamefont {A.}~\bibnamefont {Fioretti}}, \bibinfo {author} {\bibfnamefont {C.}~\bibnamefont {Gabbanini}}, \ and\ \bibinfo {author} {\bibfnamefont {G.}~\bibnamefont {Modugno}},\ }\bibfield  {title} {\enquote {\bibinfo {title} {{Evidence of superfluidity in a dipolar supersolid from nonclassical rotational inertia}},}\ }\href {\doibase 10.1126/science.aba4309} {\bibfield  {journal} {\bibinfo  {journal} {Science}\ }\textbf {\bibinfo {volume} {371}},\ \bibinfo {pages} {1162--1165} (\bibinfo {year} {2021})}\BibitemShut {NoStop}%
\bibitem [{\citenamefont {Roccuzzo}\ \emph {et~al.}(2020)\citenamefont {Roccuzzo}, \citenamefont {Gallem{\'\i}}, \citenamefont {Recati},\ and\ \citenamefont {Stringari}}]{Roccuzzo_2020}%
  \BibitemOpen
  \bibfield  {author} {\bibinfo {author} {\bibfnamefont {S.M.}\ \bibnamefont {Roccuzzo}}, \bibinfo {author} {\bibfnamefont {A.}~\bibnamefont {Gallem{\'\i}}}, \bibinfo {author} {\bibfnamefont {A.}~\bibnamefont {Recati}}, \ and\ \bibinfo {author} {\bibfnamefont {S.}~\bibnamefont {Stringari}},\ }\bibfield  {title} {\enquote {\bibinfo {title} {{Rotating a Supersolid Dipolar Gas}},}\ }\href {\doibase 10.1103/physrevlett.124.045702} {\bibfield  {journal} {\bibinfo  {journal} {Physical Review Letters}\ }\textbf {\bibinfo {volume} {124}},\ \bibinfo {pages} {045702} (\bibinfo {year} {2020})}\BibitemShut {NoStop}%
\bibitem [{\citenamefont {Roccuzzo}\ \emph {et~al.}(2022)\citenamefont {Roccuzzo}, \citenamefont {Recati},\ and\ \citenamefont {Stringari}}]{Roccuzzo_2022}%
  \BibitemOpen
  \bibfield  {author} {\bibinfo {author} {\bibfnamefont {S.~M.}\ \bibnamefont {Roccuzzo}}, \bibinfo {author} {\bibfnamefont {A.}~\bibnamefont {Recati}}, \ and\ \bibinfo {author} {\bibfnamefont {S.}~\bibnamefont {Stringari}},\ }\bibfield  {title} {\enquote {\bibinfo {title} {{Moment of inertia and dynamical rotational response of a supersolid dipolar gas}},}\ }\href {\doibase 10.1103/physreva.105.023316} {\bibfield  {journal} {\bibinfo  {journal} {Physical Review A}\ }\textbf {\bibinfo {volume} {105}},\ \bibinfo {pages} {023316} (\bibinfo {year} {2022})}\BibitemShut {NoStop}%
\bibitem [{\citenamefont {Trabucco}\ \emph {et~al.}(2026)\citenamefont {Trabucco}, \citenamefont {Poli}, \citenamefont {Mannarelli},\ and\ \citenamefont {Ferlaino}}]{Trabucco_2026}%
  \BibitemOpen
  \bibfield  {author} {\bibinfo {author} {\bibfnamefont {Silvia}\ \bibnamefont {Trabucco}}, \bibinfo {author} {\bibfnamefont {Elena}\ \bibnamefont {Poli}}, \bibinfo {author} {\bibfnamefont {Massimo}\ \bibnamefont {Mannarelli}}, \ and\ \bibinfo {author} {\bibfnamefont {Francesca}\ \bibnamefont {Ferlaino}},\ }\href@noop {} {\enquote {\bibinfo {title} {{Spontaneous counterflow in rotating supersolids}},}\ } (\bibinfo {year} {2026}),\ \Eprint {http://arxiv.org/abs/2609.07405} {arXiv:2609.07405 [cond-mat.quant-gas]} \BibitemShut {NoStop}%
\bibitem [{\citenamefont {Ramanathan}\ \emph {et~al.}(2011)\citenamefont {Ramanathan}, \citenamefont {Wright}, \citenamefont {Muniz}, \citenamefont {Zelan}, \citenamefont {Hill}, \citenamefont {Lobb}, \citenamefont {Helmerson}, \citenamefont {Phillips},\ and\ \citenamefont {Campbell}}]{Ramanathan_2011}%
  \BibitemOpen
  \bibfield  {author} {\bibinfo {author} {\bibfnamefont {A.}~\bibnamefont {Ramanathan}}, \bibinfo {author} {\bibfnamefont {K.~C.}\ \bibnamefont {Wright}}, \bibinfo {author} {\bibfnamefont {S.~R.}\ \bibnamefont {Muniz}}, \bibinfo {author} {\bibfnamefont {M.}~\bibnamefont {Zelan}}, \bibinfo {author} {\bibfnamefont {W.~T.}\ \bibnamefont {Hill}}, \bibinfo {author} {\bibfnamefont {C.~J.}\ \bibnamefont {Lobb}}, \bibinfo {author} {\bibfnamefont {K.}~\bibnamefont {Helmerson}}, \bibinfo {author} {\bibfnamefont {W.~D.}\ \bibnamefont {Phillips}}, \ and\ \bibinfo {author} {\bibfnamefont {G.~K.}\ \bibnamefont {Campbell}},\ }\bibfield  {title} {\enquote {\bibinfo {title} {{Superflow in a Toroidal Bose-Einstein Condensate: An Atom Circuit with a Tunable Weak Link}},}\ }\href {\doibase 10.1103/physrevlett.106.130401} {\bibfield  {journal} {\bibinfo  {journal} {Physical Review Letters}\ }\textbf {\bibinfo {volume} {106}},\ \bibinfo {pages} {130401} (\bibinfo {year} {2011})}\BibitemShut {NoStop}%
\bibitem [{\citenamefont {Pace}\ \emph {et~al.}(2022)\citenamefont {Pace}, \citenamefont {Xhani}, \citenamefont {Falconi}, \citenamefont {Fedrizzi}, \citenamefont {Grani}, \citenamefont {Rajkov}, \citenamefont {Inguscio}, \citenamefont {Scazza}, \citenamefont {Kwon},\ and\ \citenamefont {Roati}}]{DelPace_2022}%
  \BibitemOpen
  \bibfield  {author} {\bibinfo {author} {\bibfnamefont {G.~Del}\ \bibnamefont {Pace}}, \bibinfo {author} {\bibfnamefont {K.}~\bibnamefont {Xhani}}, \bibinfo {author} {\bibfnamefont {A.~Muzi}\ \bibnamefont {Falconi}}, \bibinfo {author} {\bibfnamefont {M.}~\bibnamefont {Fedrizzi}}, \bibinfo {author} {\bibfnamefont {N.}~\bibnamefont {Grani}}, \bibinfo {author} {\bibfnamefont {D.~Hernandez}\ \bibnamefont {Rajkov}}, \bibinfo {author} {\bibfnamefont {M.}~\bibnamefont {Inguscio}}, \bibinfo {author} {\bibfnamefont {F.}~\bibnamefont {Scazza}}, \bibinfo {author} {\bibfnamefont {W.J.}\ \bibnamefont {Kwon}}, \ and\ \bibinfo {author} {\bibfnamefont {G.}~\bibnamefont {Roati}},\ }\bibfield  {title} {\enquote {\bibinfo {title} {{Imprinting Persistent Currents in Tunable Fermionic Rings}},}\ }\href {\doibase 10.1103/physrevx.12.041037} {\bibfield  {journal} {\bibinfo  {journal} {Physical Review X}\ }\textbf {\bibinfo {volume} {12}},\ \bibinfo {pages} {041037} (\bibinfo {year} {2022})}\BibitemShut {NoStop}%
\bibitem [{\citenamefont {Xhani}\ \emph {et~al.}(2025)\citenamefont {Xhani}, \citenamefont {Barresi}, \citenamefont {Tylutki}, \citenamefont {Wlaz\l{}owski},\ and\ \citenamefont {Magierski}}]{Xhani_2025}%
  \BibitemOpen
  \bibfield  {author} {\bibinfo {author} {\bibfnamefont {Klejdja}\ \bibnamefont {Xhani}}, \bibinfo {author} {\bibfnamefont {Andrea}\ \bibnamefont {Barresi}}, \bibinfo {author} {\bibfnamefont {Marek}\ \bibnamefont {Tylutki}}, \bibinfo {author} {\bibfnamefont {Gabriel}\ \bibnamefont {Wlaz\l{}owski}}, \ and\ \bibinfo {author} {\bibfnamefont {Piotr}\ \bibnamefont {Magierski}},\ }\bibfield  {title} {\enquote {\bibinfo {title} {{Stability of persistent currents in superfluid fermionic rings}},}\ }\href {\doibase 10.1103/PhysRevResearch.7.013225} {\bibfield  {journal} {\bibinfo  {journal} {Physical Review Research}\ }\textbf {\bibinfo {volume} {7}},\ \bibinfo {pages} {013225} (\bibinfo {year} {2025})}\BibitemShut {NoStop}%
\bibitem [{\citenamefont {Schubert}\ \emph {et~al.}(2026)\citenamefont {Schubert}, \citenamefont {Mukherjee}, \citenamefont {St{\"u}rmer},\ and\ \citenamefont {Reimann}}]{Schubert2026}%
  \BibitemOpen
  \bibfield  {author} {\bibinfo {author} {\bibfnamefont {M.}~\bibnamefont {Schubert}}, \bibinfo {author} {\bibfnamefont {K.}~\bibnamefont {Mukherjee}}, \bibinfo {author} {\bibfnamefont {P.}~\bibnamefont {St{\"u}rmer}}, \ and\ \bibinfo {author} {\bibfnamefont {S.~M.}\ \bibnamefont {Reimann}},\ }\bibfield  {title} {\enquote {\bibinfo {title} {{Vorticity-Crystalline Order Coupling in Supersolids: Excitations and Reentrant Phases}},}\ }\href {\doibase 10.1103/gqlr-bgj8} {\bibfield  {journal} {\bibinfo  {journal} {Physical Review Letters}\ }\textbf {\bibinfo {volume} {136}},\ \bibinfo {pages} {183401} (\bibinfo {year} {2026})}\BibitemShut {NoStop}%
\bibitem [{\citenamefont {Tengstrand}\ \emph {et~al.}(2023)\citenamefont {Tengstrand}, \citenamefont {St\"urmer}, \citenamefont {Ribbing},\ and\ \citenamefont {Reimann}}]{Tengstrand_2023}%
  \BibitemOpen
  \bibfield  {author} {\bibinfo {author} {\bibfnamefont {M.~Nilsson}\ \bibnamefont {Tengstrand}}, \bibinfo {author} {\bibfnamefont {P.}~\bibnamefont {St\"urmer}}, \bibinfo {author} {\bibfnamefont {J.}~\bibnamefont {Ribbing}}, \ and\ \bibinfo {author} {\bibfnamefont {S.~M.}\ \bibnamefont {Reimann}},\ }\bibfield  {title} {\enquote {\bibinfo {title} {{Toroidal dipolar supersolid with a rotating weak link}},}\ }\href {\doibase 10.1103/PhysRevA.107.063316} {\bibfield  {journal} {\bibinfo  {journal} {Physical Review A}\ }\textbf {\bibinfo {volume} {107}},\ \bibinfo {pages} {063316} (\bibinfo {year} {2023})}\BibitemShut {NoStop}%
\bibitem [{\citenamefont {Mukherjee}\ \emph {et~al.}(2025)\citenamefont {Mukherjee}, \citenamefont {Cardinale},\ and\ \citenamefont {Reimann}}]{Mukherjee_2025}%
  \BibitemOpen
  \bibfield  {author} {\bibinfo {author} {\bibfnamefont {K.}~\bibnamefont {Mukherjee}}, \bibinfo {author} {\bibfnamefont {T.~Arnone}\ \bibnamefont {Cardinale}}, \ and\ \bibinfo {author} {\bibfnamefont {S.~M.}\ \bibnamefont {Reimann}},\ }\bibfield  {title} {\enquote {\bibinfo {title} {{Selective rotation and attractive persistent currents in antidipolar ring supersolids}},}\ }\href {\doibase 10.1103/PhysRevA.111.033304} {\bibfield  {journal} {\bibinfo  {journal} {Physical Review A}\ }\textbf {\bibinfo {volume} {111}},\ \bibinfo {pages} {033304} (\bibinfo {year} {2025})}\BibitemShut {NoStop}%
\bibitem [{\citenamefont {Preti}\ \emph {et~al.}(2026)\citenamefont {Preti}, \citenamefont {Antolini}, \citenamefont {Drevon}, \citenamefont {Lombardi}, \citenamefont {Fioretti}, \citenamefont {Gabbanini}, \citenamefont {Ferioli}, \citenamefont {Modugno},\ and\ \citenamefont {Biagioni}}]{Preti_2026}%
  \BibitemOpen
  \bibfield  {author} {\bibinfo {author} {\bibfnamefont {N.}~\bibnamefont {Preti}}, \bibinfo {author} {\bibfnamefont {N.}~\bibnamefont {Antolini}}, \bibinfo {author} {\bibfnamefont {C.}~\bibnamefont {Drevon}}, \bibinfo {author} {\bibfnamefont {P.}~\bibnamefont {Lombardi}}, \bibinfo {author} {\bibfnamefont {A.}~\bibnamefont {Fioretti}}, \bibinfo {author} {\bibfnamefont {C.}~\bibnamefont {Gabbanini}}, \bibinfo {author} {\bibfnamefont {G.}~\bibnamefont {Ferioli}}, \bibinfo {author} {\bibfnamefont {G.}~\bibnamefont {Modugno}}, \ and\ \bibinfo {author} {\bibfnamefont {G.}~\bibnamefont {Biagioni}},\ }\bibfield  {title} {\enquote {\bibinfo {title} {{Single-Fluid Model for Rotating Annular Supersolids and Its Experimental Implications}},}\ }\href {\doibase 10.1103/8n5y-fyh7} {\bibfield  {journal} {\bibinfo  {journal} {Physical Review Letters}\ }\textbf {\bibinfo {volume} {136}},\ \bibinfo {pages} {036001} (\bibinfo {year} {2026})}\BibitemShut {NoStop}%
\bibitem [{\citenamefont {Yanase}(2009)}]{Yanase_2009}%
  \BibitemOpen
  \bibfield  {author} {\bibinfo {author} {\bibfnamefont {Youichi}\ \bibnamefont {Yanase}},\ }\bibfield  {title} {\enquote {\bibinfo {title} {{Angular Fulde-Ferrell-Larkin-Ovchinnikov state in cold fermion gases in a toroidal trap}},}\ }\href {\doibase 10.1103/PhysRevB.80.220510} {\bibfield  {journal} {\bibinfo  {journal} {Physical Review B}\ }\textbf {\bibinfo {volume} {80}},\ \bibinfo {pages} {220510} (\bibinfo {year} {2009})}\BibitemShut {NoStop}%
\bibitem [{\citenamefont {Yoshida}\ and\ \citenamefont {Yanase}(2011)}]{Yoshida_2011}%
  \BibitemOpen
  \bibfield  {author} {\bibinfo {author} {\bibfnamefont {Tomohiro}\ \bibnamefont {Yoshida}}\ and\ \bibinfo {author} {\bibfnamefont {Youichi}\ \bibnamefont {Yanase}},\ }\bibfield  {title} {\enquote {\bibinfo {title} {{Rotating Fulde-Ferrell-Larkin-Ovchinnikov state in cold Fermi gases}},}\ }\href {\doibase 10.1103/PhysRevA.84.063605} {\bibfield  {journal} {\bibinfo  {journal} {Physical Review A}\ }\textbf {\bibinfo {volume} {84}},\ \bibinfo {pages} {063605} (\bibinfo {year} {2011})}\BibitemShut {NoStop}%
\bibitem [{\citenamefont {Yoshida}\ and\ \citenamefont {Yanase}(2012)}]{Yoshida_2012}%
  \BibitemOpen
  \bibfield  {author} {\bibinfo {author} {\bibfnamefont {T.}~\bibnamefont {Yoshida}}\ and\ \bibinfo {author} {\bibfnamefont {Y.}~\bibnamefont {Yanase}},\ }\bibfield  {title} {\enquote {\bibinfo {title} {{Crossover from Fulde-Ferrell state to Larkin-Ovchinnikov state in cold fermion gases}},}\ }\href {\doibase 10.1088/1742-6596/400/1/012086} {\bibfield  {journal} {\bibinfo  {journal} {Journal of Physics: Conference Series}\ }\textbf {\bibinfo {volume} {400}},\ \bibinfo {pages} {012086} (\bibinfo {year} {2012})}\BibitemShut {NoStop}%
\bibitem [{\citenamefont {T\"uzemen}\ \emph {et~al.}(2026)\citenamefont {T\"uzemen}, \citenamefont {Barresi}, \citenamefont {Wlaz\l{}owski}, \citenamefont {Magierski},\ and\ \citenamefont {Xhani}}]{Tuzemen_2026}%
  \BibitemOpen
  \bibfield  {author} {\bibinfo {author} {\bibfnamefont {Bu\u{g}ra}\ \bibnamefont {T\"uzemen}}, \bibinfo {author} {\bibfnamefont {Andrea}\ \bibnamefont {Barresi}}, \bibinfo {author} {\bibfnamefont {Gabriel}\ \bibnamefont {Wlaz\l{}owski}}, \bibinfo {author} {\bibfnamefont {Piotr}\ \bibnamefont {Magierski}}, \ and\ \bibinfo {author} {\bibfnamefont {Klejdja}\ \bibnamefont {Xhani}},\ }\bibfield  {title} {\enquote {\bibinfo {title} {{Impurity-controlled vortex mobility and pair breaking in fermionic superfluid rings}},}\ }\href {\doibase 10.1103/6nhm-x2p9} {\bibfield  {journal} {\bibinfo  {journal} {Physical Review Research}\ }\textbf {\bibinfo {volume} {8}},\ \bibinfo {pages} {023204} (\bibinfo {year} {2026})}\BibitemShut {NoStop}%
\bibitem [{\citenamefont {Bulgac}\ \emph {et~al.}(2012)\citenamefont {Bulgac}, \citenamefont {Forbes},\ and\ \citenamefont {Magierski}}]{Bulgac_2012}%
  \BibitemOpen
  \bibfield  {author} {\bibinfo {author} {\bibfnamefont {Aurel}\ \bibnamefont {Bulgac}}, \bibinfo {author} {\bibfnamefont {Michael~McNeil}\ \bibnamefont {Forbes}}, \ and\ \bibinfo {author} {\bibfnamefont {Piotr}\ \bibnamefont {Magierski}},\ }\bibfield  {title} {\enquote {\bibinfo {title} {{The Unitary Fermi Gas: From Monte Carlo to Density Functionals}},}\ }in\ \href@noop {} {\emph {\bibinfo {booktitle} {The BCS-BEC Crossover and the Unitary Fermi Gas}}},\ \bibinfo {series} {Lecture Notes in Physics}, Vol.\ \bibinfo {volume} {836},\ \bibinfo {editor} {edited by\ \bibinfo {editor} {\bibfnamefont {Wilhelm}\ \bibnamefont {Zwerger}}}\ (\bibinfo  {publisher} {Springer, Berlin, Heidelberg},\ \bibinfo {year} {2012})\ pp.\ \bibinfo {pages} {305--373}\BibitemShut {NoStop}%
\bibitem [{\citenamefont {Magierski}\ \emph {et~al.}(2022)\citenamefont {Magierski}, \citenamefont {Wlaz\l{}owski}, \citenamefont {Makowski},\ and\ \citenamefont {Kobuszewski}}]{Magierski_2022}%
  \BibitemOpen
  \bibfield  {author} {\bibinfo {author} {\bibfnamefont {Piotr}\ \bibnamefont {Magierski}}, \bibinfo {author} {\bibfnamefont {Gabriel}\ \bibnamefont {Wlaz\l{}owski}}, \bibinfo {author} {\bibfnamefont {Andrzej}\ \bibnamefont {Makowski}}, \ and\ \bibinfo {author} {\bibfnamefont {Konrad}\ \bibnamefont {Kobuszewski}},\ }\bibfield  {title} {\enquote {\bibinfo {title} {{Spin-polarized vortices with reversed circulation}},}\ }\href {\doibase 10.1103/PhysRevA.106.033322} {\bibfield  {journal} {\bibinfo  {journal} {Physical Review A}\ }\textbf {\bibinfo {volume} {106}},\ \bibinfo {pages} {033322} (\bibinfo {year} {2022})}\BibitemShut {NoStop}%
\bibitem [{\citenamefont {T{\"u}zemen}\ \emph {et~al.}(2020)\citenamefont {T{\"u}zemen}, \citenamefont {Kukli{\'n}ski}, \citenamefont {Magierski},\ and\ \citenamefont {Wlaz{\l}owski}}]{Tuzemen_2020}%
  \BibitemOpen
  \bibfield  {author} {\bibinfo {author} {\bibfnamefont {B.}~\bibnamefont {T{\"u}zemen}}, \bibinfo {author} {\bibfnamefont {P.}~\bibnamefont {Kukli{\'n}ski}}, \bibinfo {author} {\bibfnamefont {P.}~\bibnamefont {Magierski}}, \ and\ \bibinfo {author} {\bibfnamefont {G.}~\bibnamefont {Wlaz{\l}owski}},\ }\bibfield  {title} {\enquote {\bibinfo {title} {{Properties of Spin-polarized Impurities---Ferrons, in the Unitary Fermi Gas}},}\ }\href {\doibase 10.5506/aphyspolb.51.595} {\bibfield  {journal} {\bibinfo  {journal} {Acta Physica Polonica B}\ }\textbf {\bibinfo {volume} {51}},\ \bibinfo {pages} {595} (\bibinfo {year} {2020})}\BibitemShut {NoStop}%
\bibitem [{\citenamefont {Magierski}\ \emph {et~al.}(2021)\citenamefont {Magierski}, \citenamefont {T{\"u}zemen},\ and\ \citenamefont {Wlaz{\l}owski}}]{Magierski_2021}%
  \BibitemOpen
  \bibfield  {author} {\bibinfo {author} {\bibfnamefont {Piotr}\ \bibnamefont {Magierski}}, \bibinfo {author} {\bibfnamefont {Bu{\u{g}}ra}\ \bibnamefont {T{\"u}zemen}}, \ and\ \bibinfo {author} {\bibfnamefont {Gabriel}\ \bibnamefont {Wlaz{\l}owski}},\ }\bibfield  {title} {\enquote {\bibinfo {title} {{Dynamics of spin-polarized impurity in ultracold Fermi gas}},}\ }\href {\doibase 10.1103/physreva.104.033304} {\bibfield  {journal} {\bibinfo  {journal} {Physical Review A}\ }\textbf {\bibinfo {volume} {104}},\ \bibinfo {pages} {033304} (\bibinfo {year} {2021})}\BibitemShut {NoStop}%
\bibitem [{\citenamefont {Andreev}\ and\ \citenamefont {Lifshitz}(1969)}]{Andreev_1969}%
  \BibitemOpen
  \bibfield  {author} {\bibinfo {author} {\bibfnamefont {A.~F.}\ \bibnamefont {Andreev}}\ and\ \bibinfo {author} {\bibfnamefont {I.~M.}\ \bibnamefont {Lifshitz}},\ }\bibfield  {title} {\enquote {\bibinfo {title} {{Quantum theory of defects in crystals}},}\ }\href@noop {} {\bibfield  {journal} {\bibinfo  {journal} {Zh. Eksp. Teor. Fiz.}\ }\textbf {\bibinfo {volume} {56}},\ \bibinfo {pages} {2057} (\bibinfo {year} {1969})},\ \bibinfo {note} {[Sov. Phys. JETP {\bf 29}, 1107 (1969)]}\BibitemShut {NoStop}%
\bibitem [{\citenamefont {Chester}(1970)}]{Chester_1970}%
  \BibitemOpen
  \bibfield  {author} {\bibinfo {author} {\bibfnamefont {G.~V.}\ \bibnamefont {Chester}},\ }\bibfield  {title} {\enquote {\bibinfo {title} {{Speculations on Bose-Einstein Condensation and Quantum Crystals}},}\ }\href {\doibase 10.1103/physreva.2.256} {\bibfield  {journal} {\bibinfo  {journal} {Physical Review A}\ }\textbf {\bibinfo {volume} {2}},\ \bibinfo {pages} {256--258} (\bibinfo {year} {1970})}\BibitemShut {NoStop}%
\bibitem [{\citenamefont {Leggett}(1970)}]{Leggett_1970}%
  \BibitemOpen
  \bibfield  {author} {\bibinfo {author} {\bibfnamefont {A.~J.}\ \bibnamefont {Leggett}},\ }\bibfield  {title} {\enquote {\bibinfo {title} {{Can a Solid Be "Superfluid"?}}}\ }\href {\doibase 10.1103/physrevlett.25.1543} {\bibfield  {journal} {\bibinfo  {journal} {Physical Review Letters}\ }\textbf {\bibinfo {volume} {25}},\ \bibinfo {pages} {1543--1546} (\bibinfo {year} {1970})}\BibitemShut {NoStop}%
\bibitem [{\citenamefont {Boninsegni}\ and\ \citenamefont {Prokof'ev}(2012)}]{Boninsegni_2012}%
  \BibitemOpen
  \bibfield  {author} {\bibinfo {author} {\bibfnamefont {Massimo}\ \bibnamefont {Boninsegni}}\ and\ \bibinfo {author} {\bibfnamefont {Nikolay~V.}\ \bibnamefont {Prokof'ev}},\ }\bibfield  {title} {\enquote {\bibinfo {title} {{Colloquium: Supersolids: What and where are they?}}}\ }\href {\doibase 10.1103/revmodphys.84.759} {\bibfield  {journal} {\bibinfo  {journal} {Reviews of Modern Physics}\ }\textbf {\bibinfo {volume} {84}},\ \bibinfo {pages} {759--776} (\bibinfo {year} {2012})}\BibitemShut {NoStop}%
\bibitem [{\citenamefont {Pezz{\`e}}\ \emph {et~al.}(2024)\citenamefont {Pezz{\`e}}, \citenamefont {Xhani}, \citenamefont {Daix}, \citenamefont {Grani}, \citenamefont {Donelli}, \citenamefont {Scazza}, \citenamefont {Hernandez-Rajkov}, \citenamefont {Kwon}, \citenamefont {Del~Pace},\ and\ \citenamefont {Roati}}]{Pezze2024}%
  \BibitemOpen
  \bibfield  {author} {\bibinfo {author} {\bibfnamefont {Luca}\ \bibnamefont {Pezz{\`e}}}, \bibinfo {author} {\bibfnamefont {Klejdja}\ \bibnamefont {Xhani}}, \bibinfo {author} {\bibfnamefont {Cyprien}\ \bibnamefont {Daix}}, \bibinfo {author} {\bibfnamefont {Nicola}\ \bibnamefont {Grani}}, \bibinfo {author} {\bibfnamefont {Beatrice}\ \bibnamefont {Donelli}}, \bibinfo {author} {\bibfnamefont {Francesco}\ \bibnamefont {Scazza}}, \bibinfo {author} {\bibfnamefont {Diego}\ \bibnamefont {Hernandez-Rajkov}}, \bibinfo {author} {\bibfnamefont {Woo~Jin}\ \bibnamefont {Kwon}}, \bibinfo {author} {\bibfnamefont {Giulia}\ \bibnamefont {Del~Pace}}, \ and\ \bibinfo {author} {\bibfnamefont {Giacomo}\ \bibnamefont {Roati}},\ }\bibfield  {title} {\enquote {\bibinfo {title} {{Stabilizing persistent currents in an atomtronic Josephson junction necklace}},}\ }\href {\doibase 10.1038/s41467-024-47759-7} {\bibfield  {journal} {\bibinfo  {journal} {Nature Communications}\ }\textbf {\bibinfo {volume} {15}},\ \bibinfo {pages} {4831}
  (\bibinfo {year} {2024})}\BibitemShut {NoStop}%
\bibitem [{\citenamefont {Stetcu}\ \emph {et~al.}(2011)\citenamefont {Stetcu}, \citenamefont {Bulgac}, \citenamefont {Magierski},\ and\ \citenamefont {Roche}}]{Stetcu_2011}%
  \BibitemOpen
  \bibfield  {author} {\bibinfo {author} {\bibfnamefont {I.}~\bibnamefont {Stetcu}}, \bibinfo {author} {\bibfnamefont {A.}~\bibnamefont {Bulgac}}, \bibinfo {author} {\bibfnamefont {P.}~\bibnamefont {Magierski}}, \ and\ \bibinfo {author} {\bibfnamefont {K.~J.}\ \bibnamefont {Roche}},\ }\bibfield  {title} {\enquote {\bibinfo {title} {{Isovector giant dipole resonance from the 3D time-dependent density functional theory for superfluid nuclei}},}\ }\href {\doibase 10.1103/physrevc.84.051309} {\bibfield  {journal} {\bibinfo  {journal} {Physical Review C}\ }\textbf {\bibinfo {volume} {84}},\ \bibinfo {pages} {051309} (\bibinfo {year} {2011})}\BibitemShut {NoStop}%
\bibitem [{\citenamefont {\v{S}indik}\ \emph {et~al.}(2024)\citenamefont {\v{S}indik}, \citenamefont {Zawi\'slak}, \citenamefont {Recati},\ and\ \citenamefont {Stringari}}]{Sindik_2024}%
  \BibitemOpen
  \bibfield  {author} {\bibinfo {author} {\bibfnamefont {Marija}\ \bibnamefont {\v{S}indik}}, \bibinfo {author} {\bibfnamefont {Tomasz}\ \bibnamefont {Zawi\'slak}}, \bibinfo {author} {\bibfnamefont {Alessio}\ \bibnamefont {Recati}}, \ and\ \bibinfo {author} {\bibfnamefont {Sandro}\ \bibnamefont {Stringari}},\ }\bibfield  {title} {\enquote {\bibinfo {title} {{Sound, Superfluidity, and Layer Compressibility in a Ring Dipolar Supersolid}},}\ }\href {\doibase 10.1103/PhysRevLett.132.146001} {\bibfield  {journal} {\bibinfo  {journal} {Physical Review Letters}\ }\textbf {\bibinfo {volume} {132}},\ \bibinfo {pages} {146001} (\bibinfo {year} {2024})}\BibitemShut {NoStop}%
\bibitem [{\citenamefont {Hertkorn}\ \emph {et~al.}(2024)\citenamefont {Hertkorn}, \citenamefont {St\"urmer}, \citenamefont {Mukherjee}, \citenamefont {Ng}, \citenamefont {Uerlings}, \citenamefont {Hellstern}, \citenamefont {Lavoine}, \citenamefont {Reimann}, \citenamefont {Pfau},\ and\ \citenamefont {Klemt}}]{Hertkorn_2024}%
  \BibitemOpen
  \bibfield  {author} {\bibinfo {author} {\bibfnamefont {Jens}\ \bibnamefont {Hertkorn}}, \bibinfo {author} {\bibfnamefont {Philipp}\ \bibnamefont {St\"urmer}}, \bibinfo {author} {\bibfnamefont {Koushik}\ \bibnamefont {Mukherjee}}, \bibinfo {author} {\bibfnamefont {Kevin S.~H.}\ \bibnamefont {Ng}}, \bibinfo {author} {\bibfnamefont {Paul}\ \bibnamefont {Uerlings}}, \bibinfo {author} {\bibfnamefont {Fiona}\ \bibnamefont {Hellstern}}, \bibinfo {author} {\bibfnamefont {Lucas}\ \bibnamefont {Lavoine}}, \bibinfo {author} {\bibfnamefont {Stephanie~M.}\ \bibnamefont {Reimann}}, \bibinfo {author} {\bibfnamefont {Tilman}\ \bibnamefont {Pfau}}, \ and\ \bibinfo {author} {\bibfnamefont {Ralf}\ \bibnamefont {Klemt}},\ }\bibfield  {title} {\enquote {\bibinfo {title} {{Decoupled sound and amplitude modes in trapped dipolar supersolids}},}\ }\href {\doibase 10.1103/PhysRevResearch.6.L042056} {\bibfield  {journal} {\bibinfo  {journal} {Physical Review Research}\ }\textbf {\bibinfo {volume} {6}},\ \bibinfo {pages} {L042056}
  (\bibinfo {year} {2024})}\BibitemShut {NoStop}%
\bibitem [{\citenamefont {Ashcroft}\ and\ \citenamefont {Mermin}(1976)}]{Ashcroft_1976}%
  \BibitemOpen
  \bibfield  {author} {\bibinfo {author} {\bibfnamefont {Neil~W.}\ \bibnamefont {Ashcroft}}\ and\ \bibinfo {author} {\bibfnamefont {N.~David}\ \bibnamefont {Mermin}},\ }\href@noop {} {\emph {\bibinfo {title} {{Solid State Physics}}}}\ (\bibinfo  {publisher} {Holt, Rinehart and Winston},\ \bibinfo {address} {New York},\ \bibinfo {year} {1976})\BibitemShut {NoStop}%
\bibitem [{\citenamefont {Minakova}\ \emph {et~al.}(2026)\citenamefont {Minakova}, \citenamefont {Paiva}, \citenamefont {Frenzel}, \citenamefont {Spencer}, \citenamefont {Urban}, \citenamefont {Ringkamp}, \citenamefont {Wolf}, \citenamefont {Mussler}, \citenamefont {Juraschek},\ and\ \citenamefont {Maehrlein}}]{Minakova2026}%
  \BibitemOpen
  \bibfield  {author} {\bibinfo {author} {\bibfnamefont {Olga}\ \bibnamefont {Minakova}}, \bibinfo {author} {\bibfnamefont {Carolina}\ \bibnamefont {Paiva}}, \bibinfo {author} {\bibfnamefont {Maximilian}\ \bibnamefont {Frenzel}}, \bibinfo {author} {\bibfnamefont {Michael~S.}\ \bibnamefont {Spencer}}, \bibinfo {author} {\bibfnamefont {Joanna~M.}\ \bibnamefont {Urban}}, \bibinfo {author} {\bibfnamefont {Christoph}\ \bibnamefont {Ringkamp}}, \bibinfo {author} {\bibfnamefont {Martin}\ \bibnamefont {Wolf}}, \bibinfo {author} {\bibfnamefont {Gregor}\ \bibnamefont {Mussler}}, \bibinfo {author} {\bibfnamefont {Dominik~M.}\ \bibnamefont {Juraschek}}, \ and\ \bibinfo {author} {\bibfnamefont {Sebastian~F.}\ \bibnamefont {Maehrlein}},\ }\bibfield  {title} {\enquote {\bibinfo {title} {{Observation of angular momentum transfer among crystal lattice modes}},}\ }\href {\doibase 10.1038/s41567-026-03274-8} {\bibfield  {journal} {\bibinfo  {journal} {Nature Physics}\ }\textbf {\bibinfo {volume} {22}},\ \bibinfo {pages}
  {1100--1106} (\bibinfo {year} {2026})}\BibitemShut {NoStop}%
\bibitem [{\citenamefont {Xhani}\ \emph {et~al.}(2026)\citenamefont {Xhani}, \citenamefont {Del~Pace}, \citenamefont {Grani}, \citenamefont {Hern{\'a}ndez-Rajkov}, \citenamefont {Donelli}, \citenamefont {Roati},\ and\ \citenamefont {Pezz{\`e}}}]{xhaniPRA}%
  \BibitemOpen
  \bibfield  {author} {\bibinfo {author} {\bibfnamefont {K.}~\bibnamefont {Xhani}}, \bibinfo {author} {\bibfnamefont {G.}~\bibnamefont {Del~Pace}}, \bibinfo {author} {\bibfnamefont {N.}~\bibnamefont {Grani}}, \bibinfo {author} {\bibfnamefont {D.}~\bibnamefont {Hern{\'a}ndez-Rajkov}}, \bibinfo {author} {\bibfnamefont {B.}~\bibnamefont {Donelli}}, \bibinfo {author} {\bibfnamefont {G.}~\bibnamefont {Roati}}, \ and\ \bibinfo {author} {\bibfnamefont {L.}~\bibnamefont {Pezz{\`e}}},\ }\bibfield  {title} {\enquote {\bibinfo {title} {Tuning the critical current in toroidal superfluids via controllable impurities},}\ }\href {\doibase 10.1103/316q-596d} {\bibfield  {journal} {\bibinfo  {journal} {Phys. Rev. A}\ }\textbf {\bibinfo {volume} {113}},\ \bibinfo {pages} {L051301} (\bibinfo {year} {2026})}\BibitemShut {NoStop}%
\bibitem [{\citenamefont {Xhani}\ \emph {et~al.}(2023)\citenamefont {Xhani}, \citenamefont {Del~Pace}, \citenamefont {Scazza},\ and\ \citenamefont {Roati}}]{Xhaniatoms}%
  \BibitemOpen
  \bibfield  {author} {\bibinfo {author} {\bibfnamefont {Klejdja}\ \bibnamefont {Xhani}}, \bibinfo {author} {\bibfnamefont {Giulia}\ \bibnamefont {Del~Pace}}, \bibinfo {author} {\bibfnamefont {Francesco}\ \bibnamefont {Scazza}}, \ and\ \bibinfo {author} {\bibfnamefont {Giacomo}\ \bibnamefont {Roati}},\ }\bibfield  {title} {\enquote {\bibinfo {title} {Decay of persistent currents in annular atomic superfluids},}\ }\href {\doibase 10.3390/atoms11080109} {\bibfield  {journal} {\bibinfo  {journal} {Atoms}\ }\textbf {\bibinfo {volume} {11}},\ \bibinfo {pages} {109} (\bibinfo {year} {2023})}\BibitemShut {NoStop}%
\bibitem [{\citenamefont {Behrle}\ \emph {et~al.}(2018)\citenamefont {Behrle}, \citenamefont {Harrison}, \citenamefont {Kombe}, \citenamefont {Gao}, \citenamefont {Link}, \citenamefont {Bernier}, \citenamefont {Kollath},\ and\ \citenamefont {K{\"o}hl}}]{Behrle_2018}%
  \BibitemOpen
  \bibfield  {author} {\bibinfo {author} {\bibfnamefont {A.}~\bibnamefont {Behrle}}, \bibinfo {author} {\bibfnamefont {T.}~\bibnamefont {Harrison}}, \bibinfo {author} {\bibfnamefont {J.}~\bibnamefont {Kombe}}, \bibinfo {author} {\bibfnamefont {K.}~\bibnamefont {Gao}}, \bibinfo {author} {\bibfnamefont {M.}~\bibnamefont {Link}}, \bibinfo {author} {\bibfnamefont {J.-S.}\ \bibnamefont {Bernier}}, \bibinfo {author} {\bibfnamefont {C.}~\bibnamefont {Kollath}}, \ and\ \bibinfo {author} {\bibfnamefont {M.}~\bibnamefont {K{\"o}hl}},\ }\bibfield  {title} {\enquote {\bibinfo {title} {{Higgs mode in a strongly interacting fermionic superfluid}},}\ }\href {\doibase 10.1038/s41567-018-0128-6} {\bibfield  {journal} {\bibinfo  {journal} {Nature Physics}\ }\textbf {\bibinfo {volume} {14}},\ \bibinfo {pages} {781--785} (\bibinfo {year} {2018})}\BibitemShut {NoStop}%
\bibitem [{\citenamefont {Wlaz\l{}owski}\ \emph {et~al.}(2026)\citenamefont {Wlaz\l{}owski}, \citenamefont {Magierski}, \citenamefont {Forbes},\ and\ \citenamefont {Bulgac}}]{WSLDAToolkit}%
  \BibitemOpen
  \bibfield  {author} {\bibinfo {author} {\bibfnamefont {Gabriel}\ \bibnamefont {Wlaz\l{}owski}}, \bibinfo {author} {\bibfnamefont {Piotr}\ \bibnamefont {Magierski}}, \bibinfo {author} {\bibfnamefont {Michael~McNeil}\ \bibnamefont {Forbes}}, \ and\ \bibinfo {author} {\bibfnamefont {Aurel}\ \bibnamefont {Bulgac}},\ }\href@noop {} {\enquote {\bibinfo {title} {{W-SLDA Toolkit: A simulation platform for ultracold Fermi gases}},}\ } (\bibinfo {year} {2026}),\ \Eprint {http://arxiv.org/abs/2602.08982} {arXiv:2602.08982 [cond-mat.quant-gas]} \BibitemShut {NoStop}%
\end{thebibliography}%

\section*{End Matter}
\ssec{Method.}
We describe the system with time-dependent density functional theory in the form of the asymmetric superfluid local density approximation (TD-ASLDA)~\cite{Bulgac_2012}, implemented in the W-SLDA Toolkit~\cite{WSLDAToolkit}. In units in which $\hbar$ and the atomic mass are set to one, the energy density reads
\begin{equation}
\begin{split}
\mathcal{E} &= \alpha_\uparrow(p)\frac{\tau_\uparrow}{2}
+\alpha_\downarrow(p)\frac{\tau_\downarrow}{2}+\beta(p)\,\rho^{5/3}\\
&\quad+\frac{\gamma(p)}{\rho^{1/3}}|\nu|^2
+\sum_\sigma\bigl[1-\alpha_\sigma(p)\bigr]\frac{\mathbf{j}_\sigma^2}{2\rho_\sigma},
\end{split}
\label{eq:aslda}
\end{equation}
with $\rho=\rho_\uparrow+\rho_\downarrow$. The first two terms carry polarization-dependent effective masses; the third is the normal interaction; the fourth is the pairing interaction; and the last restores Galilean invariance. Writing ${\sum_n}'$ for a sum restricted to $0<E_n<E_c$, the densities follow from the Bogoliubov quasiparticle amplitudes $\{u_{n\sigma},v_{n\sigma}\}$ as
\begin{align}
\rho_\sigma &= {\sum_n}'\,|v_{n\sigma}|^2, &
\nu &= \frac{1}{2}{\sum_n}'\left(u_{n\uparrow}v^*_{n\downarrow}-u_{n\downarrow}v^*_{n\uparrow}\right), \nonumber\\
\tau_\sigma &= {\sum_n}'\,|\nabla v_{n\sigma}|^2, &
\mathbf{j}_\sigma &= {\sum_n}'\,\mathrm{Im}\,[v_{n\sigma}\nabla v^*_{n\sigma}],
\label{eq:densities}
\end{align}
where the cutoff $E_c$ comes from the regularization of the pairing channel, and we work at zero temperature. The coefficients $\alpha_\sigma$, $\beta$ and $\gamma$ are functions of the local polarization $p=(\rho_\uparrow-\rho_\downarrow)/\rho$ and are fitted to quantum Monte Carlo results for the homogeneous spin-imbalanced unitary gas~\cite{Bulgac_2012}. At $p=0$ the functional returns the Bertsch parameter $\xi_B\simeq0.37$ and the pairing gap $\Delta/\eF\simeq0.5$.

Stationarity of the action $S=\int dt\,[\langle0(t)|i\partial_t|0(t)\rangle-E(t)]$ gives equations of motion of time-dependent Bogoliubov--de Gennes form,
\begin{equation}
i\frac{\partial}{\partial t}
\begin{pmatrix} u_{n\uparrow} \\ v_{n\downarrow}\end{pmatrix}
=\begin{pmatrix} h_\uparrow & \Delta \\ \Delta^* & -h^*_\downarrow\end{pmatrix}
\begin{pmatrix} u_{n\uparrow} \\ v_{n\downarrow}\end{pmatrix},
\label{eq:tdaslda}
\end{equation}
with $h_\sigma=-\frac{1}{2}\nabla \cdot \alpha_{\sigma}\nabla+U_\sigma+V_{\rm ext}-\frac{i}{2}\left(\bm{A}_{\sigma}\cdot\nabla+\nabla\cdot\bm{A}_{\sigma}\right)$. The $U_\sigma=\frac{\delta\mathcal{E}}{\delta \rho_{\sigma}}$, $\Delta=-\frac{\delta\mathcal{E}}{\delta\nu^*}$ and $\bm{A}_{\sigma}=\frac{\delta\mathcal{E}}{\delta\bm{j}_{\sigma}}$ are respectively the mean-field, pairing potential, and vector potential, derived as appropriate functional derivatives. The spin-reversed components follow from the symmetry relation $u_{n\uparrow}\to v^*_{n\uparrow}$, $v_{n\downarrow}\to u^*_{n\downarrow}$, $E_n\to-E_n$. The method resolves the quasiparticles individually, so we impose no modulation, assume no symmetry, and treat the in-gap Andreev states on the same footing as the condensate. The same framework has already described the generation and proliferation of quantum vortices, spin-polarized droplets and vortex cores, and the decay of persistent currents in fermionic rings~\cite{Magierski_2019, Magierski_2022, Xhani_2025, Tuzemen_2026}.

We use a $128\times128\times16$ lattice with spacing $\Delta x=\kF^{-1}$ and periodic boundary conditions, evaluate spatial derivatives spectrally by FFT, and integrate in time with a fifth-order Adams--Bashforth--Moulton predictor--corrector scheme. We impose translational symmetry along $z$, so the solutions are plane waves in that direction. The ring is formed by the radial potential
\begin{equation}
V_{\rm ext}(r)=2\eF\left[1-f\!\left(\frac{r-R_{\rm in}+d}{d}\right)+f\!\left(\frac{r-R_{\rm out}}{d}\right)\right],
\end{equation}
with $R_{\rm in}=30\,\kF^{-1}$, $R_{\rm out}=56\,\kF^{-1}$ and ramp width $d=4\,\kF^{-1}$. The switching function $f(x)=\frac{1}{2}\{1+\tanh[\tan(\pi(x-\frac{1}{2}))]\}$ rises smoothly from $0$ at $x=0$ to $1$ at $x=1$ and is held at these values outside this interval, so $V_{\rm ext}$ vanishes for $R_{\rm in}\le r\le R_{\rm out}$ and equals $2\eF$ for $r<R_{\rm in}-d$ and $r>R_{\rm out}+d$. Both the polarization and the winding number are fixed at the static stage. At every iteration of the self-consistent loop, we impose $\Delta(\mathbf{r})\to|\Delta(\mathbf{r})|\,e^{iw_0\theta}$ inside the ring, with $\theta=\arctan(y,x)$ the azimuthal angle, and let the amplitudes relax at fixed winding~\cite{Xhani_2025}. The particle numbers, $N_\uparrow=2227$ and $N_\downarrow=1485$ ($P=20\%$), are reached by adjusting the two chemical potentials. All runs use the same total particle number, $\Ntot=N_\uparrow+N_\downarrow=3712$; the polarization only sets how it is split between the spin components. The time evolution starts from the converged polarized state with the imprint removed, so the dynamics are unconstrained. The results presented here were obtained using the approximation $\bm{A}_{\sigma}\approx 0$ in the evolution Hamiltonian, which enables simulations over sufficiently long time scales. We verified the robustness of our conclusions through additional runs using the full evolution Hamiltonian and calculations with $\alpha_{\sigma}=1$, for which $\bm{A}_{\sigma}$ vanishes identically.

\ssec{Structure analysis.}
For Fig.~\ref{fig:order}, the pairing field is interpolated linearly onto circles of radius $r$ about the ring center, with about four points per lattice spacing, and $\theta$ is measured counterclockwise from the $x$ axis. The angular Fourier coefficients $c_n(r,t)=(2\pi)^{-1}\int_0^{2\pi}d\theta\,\vert{}\Delta(r,\theta,t)\vert{}\,e^{in\theta}$ are evaluated by FFT. Nodes are the minima of $\vert{}\Delta\vert{}$ along the circle from which $\vert{}\Delta\vert{}$ rises by at least $0.3\max\vert{}\Delta\vert{}$ on both sides. On each side of a node, we fit a straight line to the unwrapped phase $\arg\Delta$ between the neighboring maximum of $\vert{}\Delta\vert{}$ and the node, leaving out the quarter of this interval closest to the node. The phase jump $\Delta\Phi\in(-\pi,\pi]$ is the difference of the two lines at the node. The mean $\langle\vert{}\Delta\Phi\vert{}\rangle$ and its standard deviation are taken over the nodes on the circle $r=\Rmid$.

\ssec{Goldstone-mode analysis.}
The kicked runs restart from the $w_0=9$, $P=20\%$ state at $t\,\eF=1250$ and are then evolved for at least an additional $t\,\eF=1000$. The pulse is centered at $t_k\,\eF=30$ after the restart and has width $t_w\,\eF=1$. Its phase $\phi_0=0.065\pi$ is the angular position of a nodal line, so every drive is referenced to the same point of the crystal. We kick at $m=1$--$8$, and at $m=11$, $13$ and $14$ for the Umklapp test, each in a separate run, and start the unkicked control from the same checkpoint. The spin-density difference is interpolated onto 720 azimuthal points on ten circles with $r=32$--$52\,\kF^{-1}$ and projected onto $\cos[m(\theta-\phi_0)]$, with $\theta$ the same angle as in $\delta V_{\sigma}$, which gives $F_m(r,t)$; for the Umklapp runs, we project onto $\cos[(\nnod-m)(\theta-\phi_0)]$ instead. After subtracting the control frame by frame, we remove the pre-pulse mean of $\delta F_m$.

We subtract a slowly varying baseline from each $\delta F_m(r,t)$ and Fourier transform it over a window of length $T$, with $T\eF=965$, that starts at $(t-t_k)\eF=5$, which gives $P_r(\omega)$. A collective displacement moves all radii in phase, which we measure with the radial coherence
\begin{equation}
C(\omega)=\frac{\bigl|\sum_r P_r(\omega)\bigr|^2}{N_r\sum_r|P_r(\omega)|^2},
\end{equation}
where $N_r=10$. It equals $1$ for an in-phase oscillation. We keep only maxima of the power $\sum_r|P_r|^2$ with $C>0.40$, a threshold that uncorrelated noise rarely exceeds. A peak is discarded if the same projection of the control run carries a radially coherent peak more than ten times stronger at that frequency. In each channel, $\omega(m)$ is the strongest remaining peak; for the Umklapp runs, we instead take the remaining peak closest to the frequency of the directly kicked $\nnod-m$ mode and compare the two. The dominant feature of the $m=3$--$5$ spectra, an $m$-independent peak near $0.028\,\eF$, fails the coherence criterion: the inner and outer parts of the ring oscillate in antiphase, as in a radial mode with a node across the ring width, rather than moving together as in a crystal translation. Since the true mode frequencies are unknown, we calibrate our extraction procedure on artificial signals of known frequency. Specifically, we inject an oscillation with the measured mode's amplitude and damping into the data and assess how accurately the analysis recovers its frequency. This gives an uncertainty of at most $10^{-3}\,\eF$ for $m\leq3$, and $6\times10^{-3}$ and $8\times10^{-3}\,\eF$ for the more strongly damped $m=4$ and $5$ modes. A weighted linear fit over $m=1$--$5$, with errors scaled by the square root of the reduced $\chi^2$, gives $\omega/\eF=0.0198(11)\,m+0.004(2)$. At radius $R$ the azimuthal wavenumber is $q=m/R$, so the mode speed is $c=R\,d\omega/dm$. With $\eF=\kF v_F/2$, this gives $c=0.42\,v_F$ at $\Rmid=42\,\kF^{-1}$ and $0.32$--$0.52\,v_F$ across $r=32$--$52\,\kF^{-1}$.

\end{document}